\documentclass[aps,prb,preprint,superscriptaddress,longbibliography]{revtex4-2}
\usepackage[utf8]{inputenc}
\usepackage{amsmath,amssymb,bm}
\usepackage{graphicx}
\usepackage{xcolor}
\usepackage{array,booktabs,multirow,tabularx}
\usepackage{siunitx}
\usepackage{cases}
\usepackage{tikz}
\usepackage{comment}
\usepackage{mdframed}
\usepackage{lipsum}
\usepackage[normalem]{ulem}

\usepackage{caption}
\usepackage{subcaption}

\usepackage{xurl}
\usepackage[hidelinks]{hyperref}
\usepackage[capitalise]{cleveref}

\usepackage{microtype}
\begin{document}

\title{Phonon-Mediated Thermal Transport in Nanocrystalline Silicon Using Machine-Learning Interatomic Potentials}

\author{Houssem Rezgui}
\email{houssem.rezgui@inl.int}
\affiliation{{International Iberian Nanotechnology Laboratory (INL), Av. Mte. José Veiga s/n, 4715-330 Braga, Portugal.}}

\author{Catalina Coll Benejam}
\email{ccollbenejam@ub.edu}
\affiliation{{Departament de Física Aplicada, Universitat de Barcelona (UB), 08028 Barcelona, Spain}}
\affiliation{{LENS-MIND, Institute of Nanoscience and Nanotechnology (IN2UB), 08028 Barcelona, Spain.}}

\author{Clivia M. Sotomayor Torres}
\email{clivia.sotomayor@inl.int}
\affiliation{{International Iberian Nanotechnology Laboratory (INL), Av. Mte. José Veiga s/n, 4715-330 Braga, Portugal.}}

\author{Miguel Pruneda}
\email{mpruneda@csic.es}
\affiliation{{Catalan Institute of Nanoscience and Nanotechnology (ICN2), CSIC and BIST, Campus UAB, 08193 Bellaterra, Spain.}}
\affiliation{{Nanomaterials and Nanotechnology Research Center (CINN-CSIC), Universidad de Oviedo (UO), Principado de Asturias, 33940 El Entrego, Spain.}}

\date{\today}

\begin{abstract}
Understanding phonon-mediated heat transport in structurally complex materials remains a central challenge for next-generation electronic and nanomechanical devices, where grain boundaries and interfacial disorder strongly limit thermal dissipation. Although classical interatomic potentials enable large-scale simulations, their limited transferability can lead to inaccuracies in vibrational properties and interfacial phonon scattering. In this work, we develop a machine learning–based framework for modeling thermal transport in bulk and nanocrystalline silicon by combining Gaussian approximation potential and multi-atomic cluster expansion models with lattice-dynamical calculations and molecular dynamics. Harmonic and anharmonic force constants derived from machine-learning interatomic potentials (MLIPs) are used within a unified Phonopy/Phono3py workflow to compute phonon dispersions, lifetimes, and lattice thermal conductivity, providing an internally consistent description of vibrational properties. In nanocrystalline silicon, non-equilibrium molecular dynamics simulations directly quantify the thermal boundary resistance associated with grain boundaries and reveal its sensitivity to interfacial roughness and the underlying interatomic description. Compared with the Stillinger–Weber and Tersoff potentials, the MLIPs provide a quantitatively accurate and internally consistent description of bulk and interfacial phonon transport, enabling better predictive modeling of nanoscale thermal transport in low-dimensional materials.
\end{abstract}

\maketitle
\section{Introduction}

Thermal transport in nanoelectromechanical systems (NEMS) remains a significant challenge, particularly in quantum nanomechanical resonators where heat flow can exhibit pronounced fluctuations and nonlocal transport effects~\cite{Yang2020,Morell2019MoSe2}. In strongly confined structures, phonon scattering, interfacial interactions, and thermal energy exchange lead to spatially non-uniform temperature distributions that directly influence mechanical stability, dissipation mechanisms, and quantum coherence~\cite{Jia2026}. From a broader transport perspective, efficient thermal management has become a critical challenge for nanoscale devices, where heat dissipation is increasingly constrained by reduced dimensions, structural disorder, and interfacial phonon scattering~\cite{Cahill2014APR,Pop2010NanoRes,Chen2021NRP}. As device dimensions approach characteristic phonon length scales, classical descriptions of heat conduction based on Fourier's law progressively lose validity, and thermal transport becomes increasingly governed by phonon scattering, coherence loss, and non-equilibrium effects~\cite{Chen2021NRP,Chen2005Book,Minnich2011PRL,Guo2018PRB}. In crystalline silicon, heat is predominantly carried by phonons and can be described within a lattice-dynamical framework that incorporates phonon–phonon interactions through the Boltzmann transport equation~\cite{Broido2007APL,PhysRevE.83.056706,Protik2022,Esfarjani2011PRB,Mingo2003PRB,PhysRevLett.110.265506,Hua2017PhysRevB,Zhang2025APL,kz9s-y611}. At these length scales, thermal conduction is fundamentally determined by phonon group velocities, mode-dependent lifetimes, and anharmonic interactions~\cite{Shanks1963PhysRev,Glassbrenner1964PhysRev,Li2012PhysRevB,Regner2013NatCommun,Johnson2013,Sledzinska2020AFM,Reig2022AdvMater,Dudde2025MTP,Raciti2025AdvSci}. In nanostructured and polycrystalline silicon, internal interfaces such as grain boundaries, surfaces, and defects strongly perturb phonon propagation, resulting in reduced thermal conductivity and the emergence of interface-dominated heat transport regimes~\cite{Bodapati2006PhysRevB,Termentzidis2009PhysRevB, Gordiz2015NJP,Isotta2024AdvFunctMater}.

Grain boundaries (GBs) are ubiquitous in nanocrystalline silicon and play a central role in determining its thermal performance. These interfaces act as effective phonon scattering centers that disrupt lattice periodicity, alter vibrational spectra, and limit phonon mean free paths (MFPs), with grain-boundary transmission remaining high below approximately $2$--$3~\mathrm{THz}$ but decreasing rapidly above $3$--$4~\mathrm{THz}$~\cite{Yang2017SciRep}. Experimental and computational studies consistently report a reduction in thermal conductivity from approximately $148~\mathrm{W\,m^{-1}\,K^{-1}}$ in single-crystal silicon at room temperature to approximately $10~\mathrm{W\,m^{-1}\,K^{-1}}$ in nanocrystalline silicon with grain sizes of around $100~\mathrm{nm}$~\cite{Maire2022AdvFunctMater,Yang2017SciRep}. This reduction is primarily attributed to enhanced phonon scattering at GBs, which strongly impacts heat-carrying acoustic modes in bulk silicon~\cite{Minnich2011PRL,Henry2009PRB}. From a microscopic perspective, GBs introduce structural disorder, break translational symmetry, and modify local bonding environments, thereby reshaping the vibrational landscape. These alterations give rise to localized vibrational states and frequency-dependent phonon scattering mechanisms~\cite{Watanabe2007JApplPhys,Bodapati2006PhysRevB,Haas2023NanoLett}. A detailed understanding of phonon interactions with GBs across multiple length and time scales is therefore essential for predictive modeling and thermal engineering of silicon nanostructures.

Over the past decades, atomistic simulations have provided detailed insights into grain-boundary structures, thermodynamic stability, and defect formation in silicon~\cite{Plimpton1995JCP,Stillinger1985PRB,Tersoff1988PRB}. Molecular dynamics (MD) simulations~\cite{Maiti1997SSC,Volz1999PRB} have been extensively employed to investigate thermal transport in nanostructured materials because they naturally account for anharmonic phonon–phonon interactions and enable atomistic simulation of phonon scattering beyond continuum approximations~\cite{Volz1999PRB}. Non-equilibrium MD (NEMD) approaches have been applied to extract lattice thermal conductivity and thermal boundary resistance in systems containing interfaces and grain boundaries~\cite{Bodapati2006PhysRevB, Albrigi2024APL,Farzadian2024APL}. Despite these advances, accurately describing phonon-mediated heat transfer across interfaces remains challenging due to the coupled effects of harmonic lattice dynamics, anharmonic interactions, and interfacial disorder~\cite{Yang2023JAP}, thereby imposing stringent requirements on the underlying interatomic potential. Classical interatomic potentials, such as the Stillinger–Weber and Tersoff models, have long been employed to study thermal transport in silicon due to their computational efficiency and well-established performance in reproducing bulk structural and elastic properties~\cite{Stillinger1985PRB,Tersoff1988PRB}. However, their restricted functional forms and parameterization strategies can introduce significant inaccuracies in vibrational properties, including phonon dispersions, group velocities, and scattering rates~\cite{Broido2007APL,Esfarjani2011PRB}. These limitations become increasingly pronounced in environments that deviate from ideal crystalline symmetry, such as grain boundaries and other interfaces, where local coordination and atomic environments differ from bulk conditions. Consequently, predictions of phonon lifetimes, lattice thermal conductivity, and interfacial thermal resistance obtained using classical potentials may exhibit substantial quantitative deviations and, in some cases, altered transport trends~\cite{Fujii2022CompMaterSci,Khot2025JAP}.

In recent years, machine-learning interatomic potentials (MLIPs) have emerged as a powerful alternative for modeling complex materials with near first-principles accuracy at a substantially reduced computational cost~\cite{Corradini2025npjCompMater,Deringer2019AdvMater,Shan2025PRB,Guo2025JAP,Wang2023PhysRevB,Behler2007PRL,Bartok2010PRL,Bartok2018PRX,Zhang2018PRLDeepMD,Wang2026MTP}. By learning high-dimensional representations of the potential energy surface from density functional theory reference datasets, MLIPs capture many-body interactions, diverse local atomic environments, and anharmonic effects that are difficult to describe using fixed analytical forms. Unlike empirical models, they offer a systematic improvement and enhanced transferability across varied configurations, making them well suited for structurally heterogeneous systems. Among the most successful MLIP approaches for silicon are the Gaussian Approximation Potential (GAP) and the equivariant neural-network-based MACE framework~\cite{Bartok2010PRL,Bartok2018PRX,Batatia2022NeurIPS,Kovacs2023JCP}. The GAP model combines two-body and many-body Smooth Overlap of Atomic Positions (SOAP) descriptors to characterize local environments, enabling accurate treatment of interactions relevant to phonon transport~\cite{Bartok2013PhysRevB,Caro2019PhysRevB,Deringer2021ChemRev}, while the MACE framework employs equivariant message-passing neural networks that preserve fundamental physical symmetries and capture complex many-body correlations, providing a complementary representation of the potential energy surface~\cite{Batatia2022NeurIPS,Kovacs2023JCP}. These capabilities make MLIPs particularly attractive for simulations of phonon-mediated thermal transport, where accurate second- and higher-order force constants are required to describe both harmonic lattice dynamics and anharmonic phonon interactions~\cite{PhysRevB.102.195412}. By faithfully representing the underlying potential energy surface, MLIPs enable reliable predictions of phonon lifetimes and the resulting lattice thermal conductivity in bulk and structurally complex systems~\cite{Rajabpour2025IJTS}. Consequently, MLIPs provide a promising route toward predictive, large-scale simulations of thermal transport in nanostructured silicon, including systems containing grain boundaries and other interfaces where classical potentials often exhibit limitations~\cite{Fujii2022CompMaterSci}. Despite this progress, systematic investigations of phonon transport across grain boundaries using MLIPs remain limited, and the extent to which modern machine-learning models modify interfacial phonon scattering relative to classical potentials has not yet been fully clarified. A comprehensive understanding of interfacial thermal transport thus requires combining accurate interatomic potentials with phonon-resolved analysis techniques that directly connect microscopic vibrational properties to macroscopic transport behavior.
In this work, we investigate phonon-mediated thermal transport in bulk and nanocrystalline silicon using MLIPs in combination with lattice-dynamical analysis and non-equilibrium molecular dynamics simulations. We first benchmark GAP, MACE, Stillinger–Weber, and Tersoff potentials against reference bulk silicon data, assessing their ability to reproduce phonon dispersions, vibrational properties, and lattice thermal conductivity. Unlike the perturbed MD approach used by Fuji and Seko~\cite{Fujii2022CompMaterSci}, which is primarily suited for evaluating bulk thermal conductivity in homogeneous systems, we employ NEMD simulations to investigate the effects of varying grain-boundary roughness amplitudes and disorder on thermal transport. We then analyze symmetric tilt grain boundaries with varying misorientation angles, examining phonon lifetimes, mean free paths, and frequency-resolved transport characteristics within the grain-boundary region. Interfacial heat transfer is quantified using MLIPs-informed non-equilibrium molecular dynamics to evaluate thermal boundary resistance, thereby establishing a unified framework for assessing the predictive capability of MLIPs for interfacial thermal transport and phonon engineering in low-dimensional materials. Overall, the results demonstrate that MLIPs provide an accurate and consistent description of phonon-mediated heat transport, while highlighting the critical role of grain-boundary structure in governing phonon scattering and thermal boundary resistance. The paper is organized as follows: Section~II outlines the computational methods and the development of the interatomic potentials. Section~III presents the bulk silicon validation and benchmarking results. Section~IV focuses on phonon transport in grain-boundary systems, including vibrational properties, phonon lifetimes, and mean free paths. Section~V investigates interfacial thermal transport and thermal boundary resistance using NEMD simulations. The main conclusions are summarized in Sec.~VI.


\section{Bulk Silicon: Model Validation}
\label{sec:bulk}

A reliable description of interfacial thermal transport requires an accurate representation of bulk vibrational properties~\cite{PhysRevB.35.9120,Tadano2014JPCM, Cheng2021NatCommun}. We therefore begin by validating the interatomic potentials against the phonon and thermal transport characteristics of crystalline silicon, thereby establishing a consistent baseline for assessing the performance of our molecular-dynamics–based machine-learning models. Bulk silicon was modeled as a diamond cubic structure generated using Atomsk~\cite{Hirel2015Atomsk}. A \(3 \times 3 \times 3\) supercell of the conventional cubic unit cell, comprising 216 atoms, was constructed using the experimental lattice parameter \(a = 5.431\)~\AA. Periodic boundary conditions were applied in all directions. Molecular dynamics simulations were performed using LAMMPS~\cite{Thompson2022LAMMPS} interfaced with the QUIP/QUIPY package ~\cite{Kermode2020-wu} and employing a previously developed high-accuracy Gaussian Approximation Potential (GAP) for silicon~\cite{Bartok2018PRX}, with a time step of 1~fs. The GAP model was trained on density functional theory (DFT) data covering crystalline and amorphous configurations~\cite{Bartok2018PRX}. It describes Si–Si interactions using a many-body framework based on local atomic environments, enabling an accurate representation of both bulk and defected structures, including grain boundaries. The structure was first relaxed via energy minimization, after which the system was equilibrated at 300~K following a staged protocol: atomic velocities were initialized from a Maxwell–Boltzmann distribution, followed by equilibration in the canonical (NVT) ensemble using a Nos\'e–Hoover thermostat~\cite{Nose1984JCP,Hoover1985PRA}. The simulation cell volume was subsequently relaxed in the isothermal–isobaric (NPT) ensemble at zero external pressure, and the system was finally re-equilibrated in the NVT ensemble to ensure thermal stability. The total equilibration time amounted to 60~ps, comprising an initial NVT stage (20~ps), followed by an NPT relaxation at zero pressure (20~ps), and a final NVT equilibration step (20~ps). Atomic configurations extracted from the equilibrated trajectories, together with their corresponding energies and forces, were stored in extended XYZ (\texttt{extxyz}) format and used as the reference dataset for subsequent model training. To enable a consistent comparison of interatomic models across bulk and interfacial environments, we construct a unified dataset of energies and forces representative of the relevant configurations. Using this dataset, four interatomic potentials were considered, including a retrained GAP model (\texttt{Si\_Phonon.xml}), an equivariant neural-network model based on the MACE framework, and the empirical Stillinger--Weber (SW) and Tersoff potentials. Both GAP and MACE were trained on the same reference dataset, enabling a consistent and unbiased assessment of their predictive performance relative to the empirical models. Further details of the training procedure are provided in the Supplementary Material S1.
The GAP model (\texttt{Si\_Phonon.xml}) was optimized using the \texttt{gap\_fit} program within the QUIP framework, employing a combined two-body distance descriptor and a many-body SOAP descriptor. The SOAP representation used $n_{\max}=7$, $l_{\max}=6$, $\zeta=2$, and 800 sparse points, with energies and forces included simultaneously in the training objective, using optimized regularization parameters. The SW and Tersoff potentials were used as implemented in the QUIP/QUIPY interface,  while the MACE model was trained using the \texttt{mace\_run\_train} workflow with a validation fraction of 0.20, optimizing joint energy and force targets for up to 200 epochs on a GPU. This unified training protocol establishes a consistent baseline to evaluate vibrational and thermal transport properties. The trained models were then used to compute force constants of bulk silicon. The harmonic and anharmonic force constants of bulk silicon were subsequently computed using the finite-displacement method implemented in Phonopy and Phono3py~\cite{phonopy-phono3py-JPSJ,phonopy-phono3py-JPCM}. Atomic forces for the displaced supercells were evaluated using the trained GAP and MACE models through their respective Atomic Simulation Environment (ASE) interfaces~\cite{HjorthLarsen2017ASE}. Second- and third-order force constants were computed using a $3 \times 3 \times 3$ displacement supercell within the Phonopy/Phono3py framework, with displacement amplitudes of 0.01~\AA\ (harmonic) and 0.03~\AA\ (anharmonic). The resulting second- and third-order force-constant tensors were reconstructed from the calculated forces and stored in HDF5 format, enabling computation of phonon dispersions, phonon lifetimes, and lattice thermal transport properties. A detailed description of the training procedure and force-constant workflow is provided in the Supplementary Material (see Section~S1).

\begin{figure}[t]
    \centering
    \begin{subfigure}{0.48\linewidth}
        \centering
        \includegraphics[width=\linewidth]{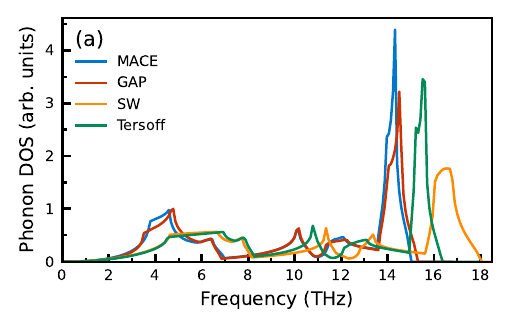}
        \label{fig:Si_phonon_a}
    \end{subfigure}
    \hfill
    \begin{subfigure}{0.48\linewidth}
        \centering
        \includegraphics[width=\linewidth]{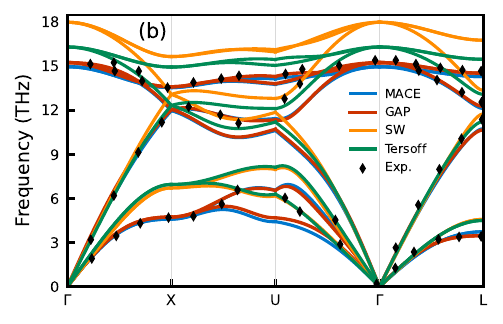}
        \label{fig:Si_phonon_b}
    \end{subfigure}

    \caption{Bulk silicon phonon properties computed using different interatomic potentials:
    (a) phonon density of states (DOS), and
    (b) phonon dispersion relations compared with experimental data from Ref.~\cite{PhysRevB.35.9120}.}
    
    \label{fig:Si_phonon}
\end{figure}

\begin{figure}[t]
    \centering
    \includegraphics[width=\linewidth]{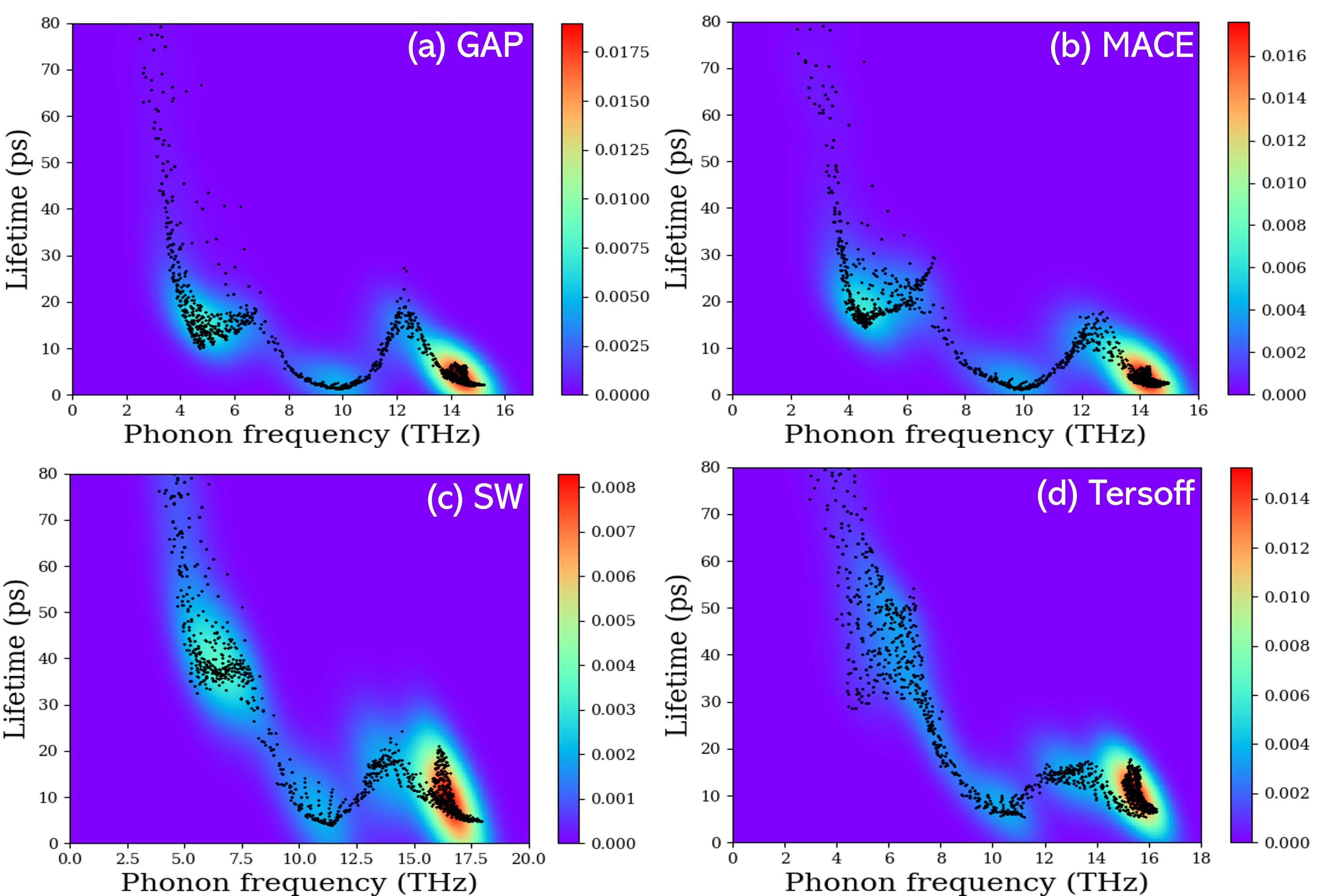}
    \caption{Phonon lifetimes of bulk silicon as a function of phonon frequency predicted by different interatomic potentials at 300~K. (a) GAP, (b) MACE, (c) Stillinger--Weber, and (d) Tersoff. The color scale indicates the probability density of phonon modes in the frequency–lifetime distribution.}
    \label{fig:Si_lifetime}
\end{figure}

\begin{figure}[t]
    \centering
    \begin{subfigure}{0.48\linewidth}
        \includegraphics[width=\linewidth]{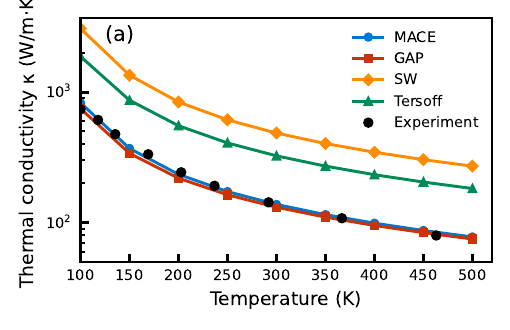}
    \end{subfigure}
    \hfill
    \begin{subfigure}{0.48\linewidth}
        \includegraphics[width=\linewidth]{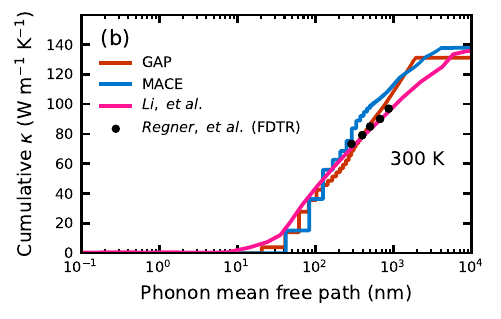}
    \end{subfigure}

    \vspace{0.6em}

    \begin{subfigure}{0.48\linewidth}
        \includegraphics[width=\linewidth]{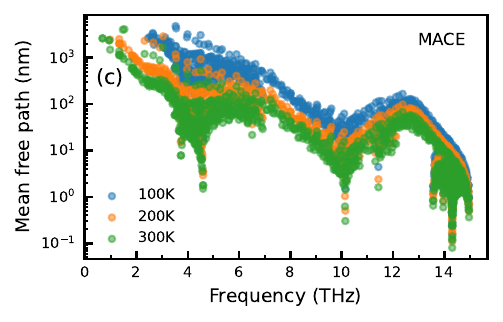}
    \end{subfigure}
    \hfill
    \begin{subfigure}{0.48\linewidth}
        \includegraphics[width=\linewidth]{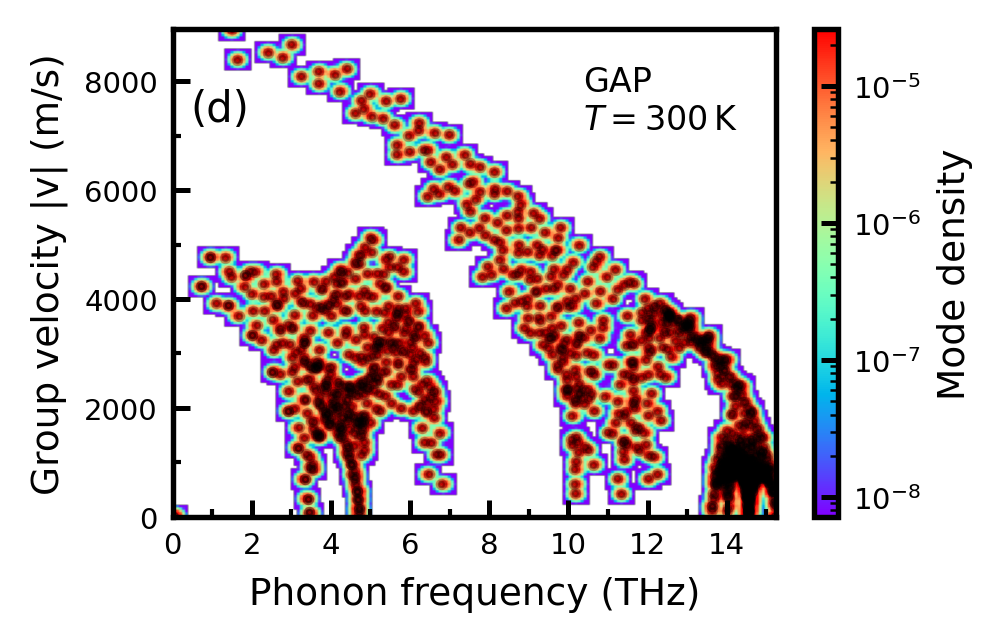}
    \end{subfigure}
    \caption{Bulk silicon thermal transport properties: (a) thermal conductivity compared with experimental data from ~\cite{Glassbrenner1964PhysRev}, (b) cumulative thermal conductivity, compared with theoretical results from ~\cite{Li2012PhysRevB} and experimental FDTR measurements from ~\cite{Regner2013NatCommun}. (c) phonon mean free path, and (d) phonon group velocity. The color scale represents the density of phonon modes.}
    \label{fig:Si_thermal}
\end{figure}

Figure~\ref{fig:Si_phonon} summarizes the vibrational properties of bulk silicon obtained using different interatomic potentials. The phonon density of states in Figure~\ref{fig:Si_phonon}(a) shows that GAP and MACE accurately reproduce the overall spectral shape and bandwidth, including the clear separation between acoustic and optical branches. In contrast, the Stillinger–Weber and Tersoff potentials exhibit noticeable deviations in the high-frequency optical region, indicating deficiencies in their harmonic force constants. The dispersion relations in Figure~\ref{fig:Si_phonon}(b) further confirm this trend, as GAP and MACE closely follow the expected behavior along high-symmetry directions and agree well with available experimental measurements~\cite{PhysRevB.35.9120}. The empirical potentials, however, systematically overestimate optical-mode frequencies and slightly distort branch curvature near the Brillouin-zone boundaries. These discrepancies alter phonon group velocities and consequently affect thermal transport predictions. Figure~\ref{fig:Si_lifetime} presents the frequency-dependent phonon lifetimes predicted by the different interatomic potentials. Figures~\ref{fig:Si_lifetime}(a) and \ref{fig:Si_lifetime}(b) show that GAP and MACE yield comparable lifetime distributions across the full frequency range. Both models exhibit physically consistent behavior, with acoustic phonons possessing finite but moderate lifetimes and a systematic reduction toward higher frequencies due to increasing anharmonic scattering. The overall magnitude and frequency scaling are similar between GAP and MACE, indicating that both MLIPs provide a consistent description of third-order force constants. In contrast, Figures~\ref{fig:Si_lifetime}(c) and \ref{fig:Si_lifetime}(d) demonstrate that the SW and Tersoff potentials predict significantly longer phonon lifetimes, particularly in the low- and mid-frequency acoustic regions. This overestimation reflects an underprediction of the anharmonic phonon–phonon scattering strength. The artificially extended lifetimes lead to unrealistically long mean free paths and directly contribute to the overestimated thermal conductivity obtained with these empirical models. Overall, the comparison confirms that the MLIPs provide a balanced and physically realistic description of anharmonic phonon scattering, whereas the empirical models systematically underestimate scattering rates. The temperature dependence of the lattice thermal conductivity shown in Figure~\ref{fig:Si_thermal}(a) demonstrates that GAP and MACE provide the most accurate description among the considered potentials. The temperature dependence of the lattice thermal conductivity shows that GAP and MACE are in good agreement with experimental data ~\cite{Glassbrenner1964PhysRev}, providing external validation of the MLIP-based description. In contrast, SW and Tersoff exhibit clear deviations, consistent with the inaccurate phonon lifetimes discussed above. Both MLIPs capture not only the correct magnitude of $\kappa$ but also its temperature dependence, including the systematic decrease with increasing temperature arising from enhanced phonon–phonon scattering. The agreement with experimental measurements confirms that the harmonic and anharmonic force constants derived from the MLIPs provide an accurate microscopic description of phonon transport in bulk silicon. In contrast, the Stillinger–Weber and Tersoff models systematically overestimate the thermal conductivity, particularly at low temperatures where long-wavelength acoustic phonons dominate heat conduction. Figure~\ref{fig:Si_thermal}(b) shows the cumulative lattice thermal conductivity as a function of phonon mean free path at 300~K. The cumulative curves obtained using GAP and MACE closely match the first-principles results reported by Li \emph{et al.}~\cite{Li2012PhysRevB} and the experimental measurements based on frequency-domain thermoreflectance (FDTR)~\cite{Regner2013NatCommun}, confirming that both MLIPs accurately reproduce the distribution of heat-carrying phonons across length scales extending to hundreds of nanometers. Figure~\ref{fig:Si_thermal}(c) presents the phonon mean free path as a function of frequency computed using the MACE model at 100~K, 200~K, and 300~K. The results exhibit the expected temperature dependence, with mean free paths systematically decreasing as temperature increases due to enhanced anharmonic scattering. Low-frequency acoustic modes possess the largest mean free paths, whereas higher-frequency modes experience progressively stronger scattering. Figure~\ref{fig:Si_thermal}(d) displays the phonon group-velocity magnitude map at 300~K obtained using the GAP model, highlighting the large velocities of low-frequency acoustic branches and the significantly smaller velocities of optical modes. The internal consistency among phonon dispersion, lifetimes, mean free paths, group velocities, and macroscopic thermal conductivity demonstrates that the MLIP-based framework provides a quantitatively reliable and physically coherent description of phonon transport in bulk silicon.
Taken together, these results show that the MLIPs achieve accuracy approaching first-principles predictions for both harmonic and anharmonic phonon properties. The agreement with experimental and theoretical data supports the reliability of the MLIP-derived force constants while retaining the computational efficiency required for large-scale simulations. These validated bulk results provide the basis for the subsequent nanoscale thermal transport analysis.


\section{Nanocrystalline Silicon: Grain Boundary Model}

\begin{figure}[t]
\centering

\begin{subfigure}{0.48\linewidth}
    \centering
    \includegraphics[width=\linewidth]{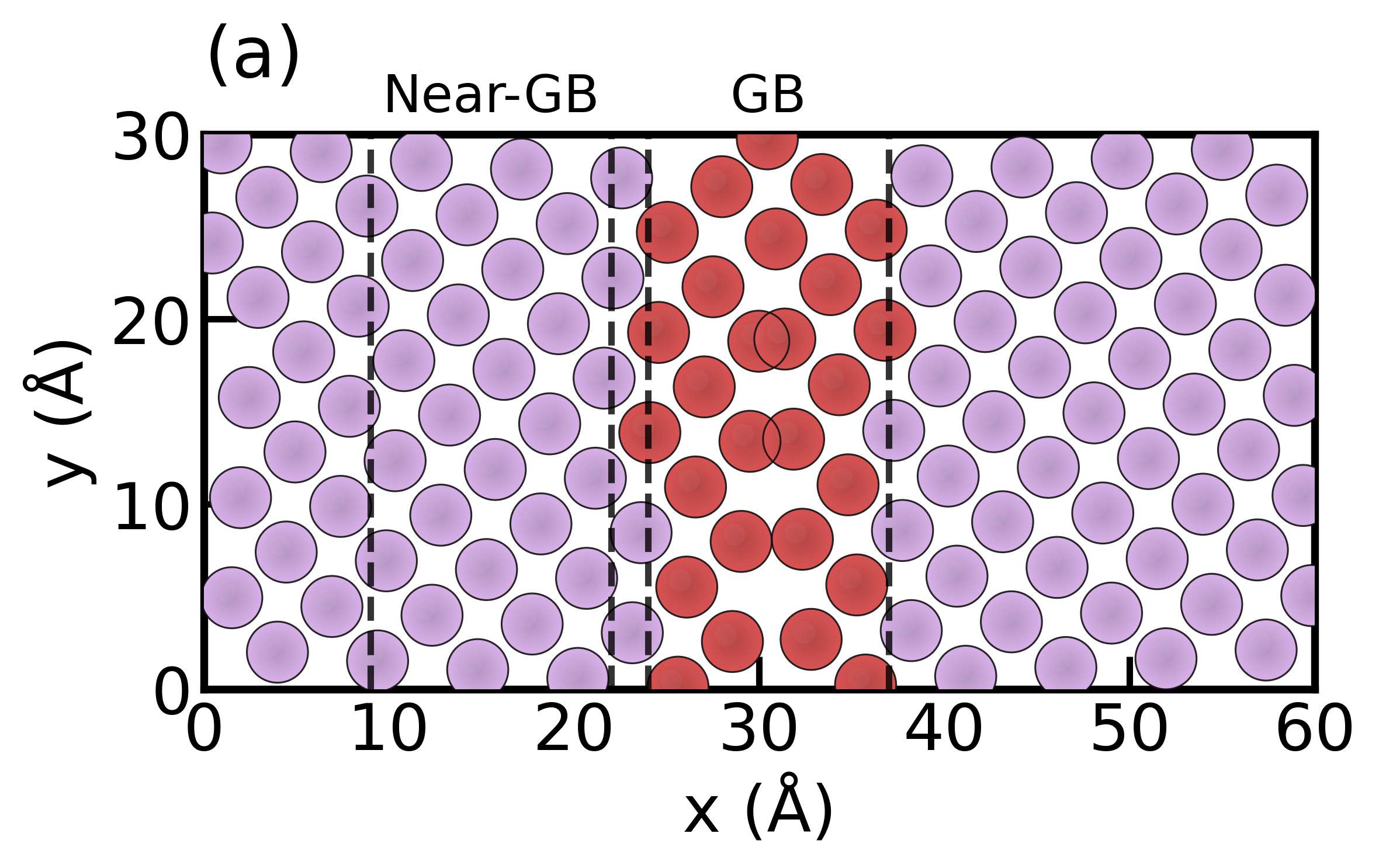}
\end{subfigure}
\hfill
\begin{subfigure}{0.48\linewidth}
    \centering
    \includegraphics[width=\linewidth]{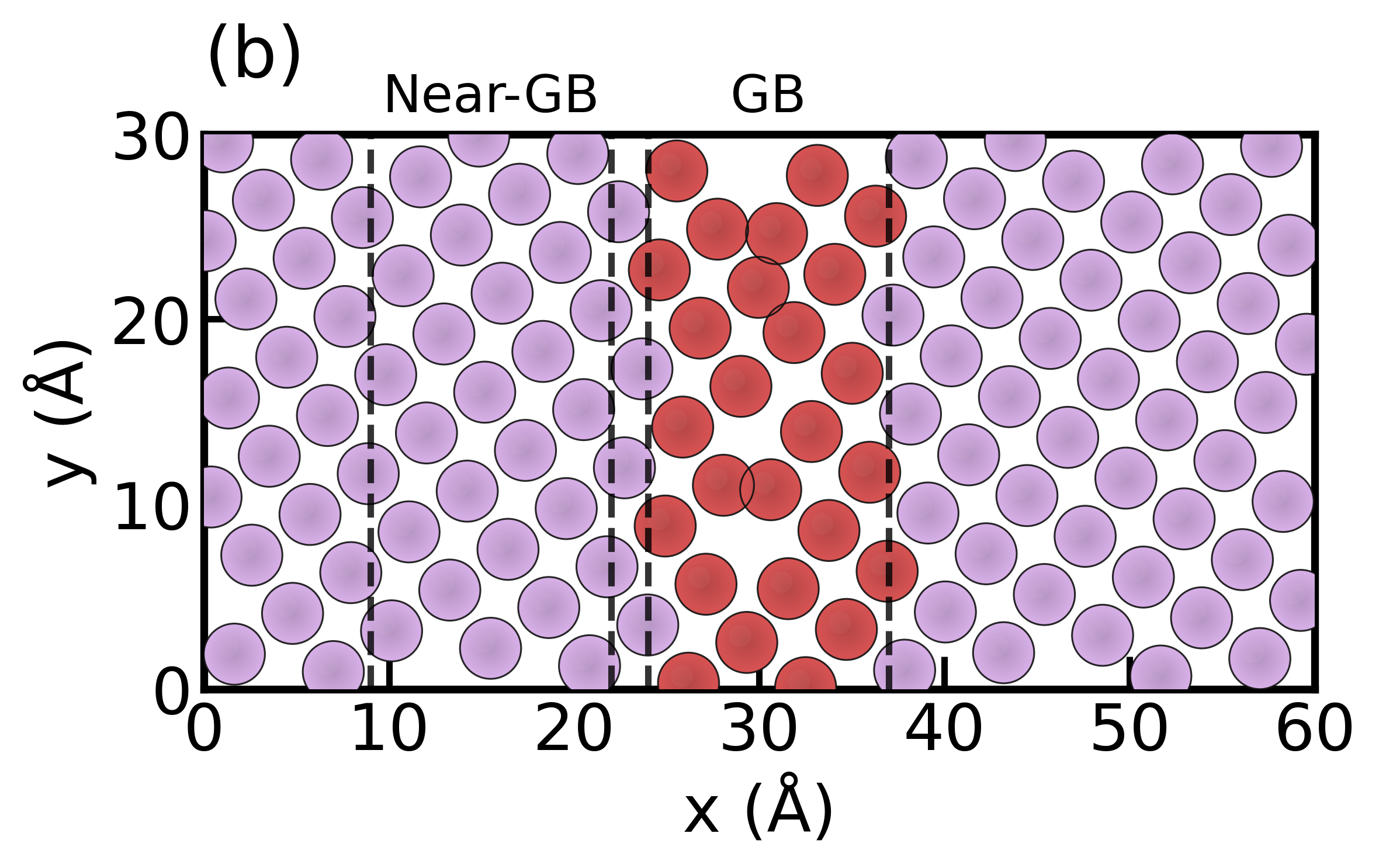}
\end{subfigure}

\vspace{0.6em}

\begin{subfigure}{0.48\linewidth}
    \centering
    \includegraphics[width=\linewidth]{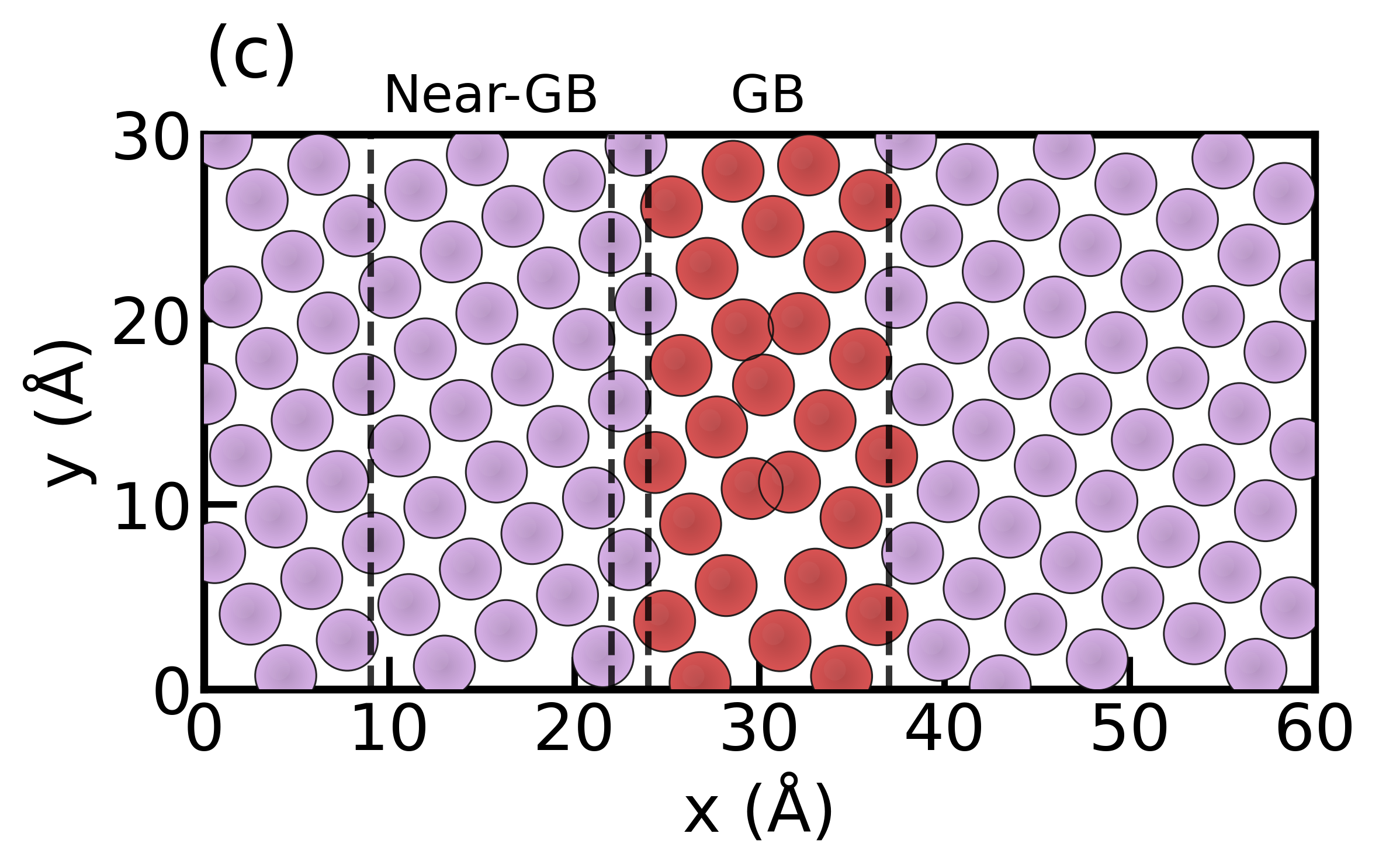}
\end{subfigure}
\hfill
\begin{subfigure}{0.48\linewidth}
    \centering
    \includegraphics[width=\linewidth]{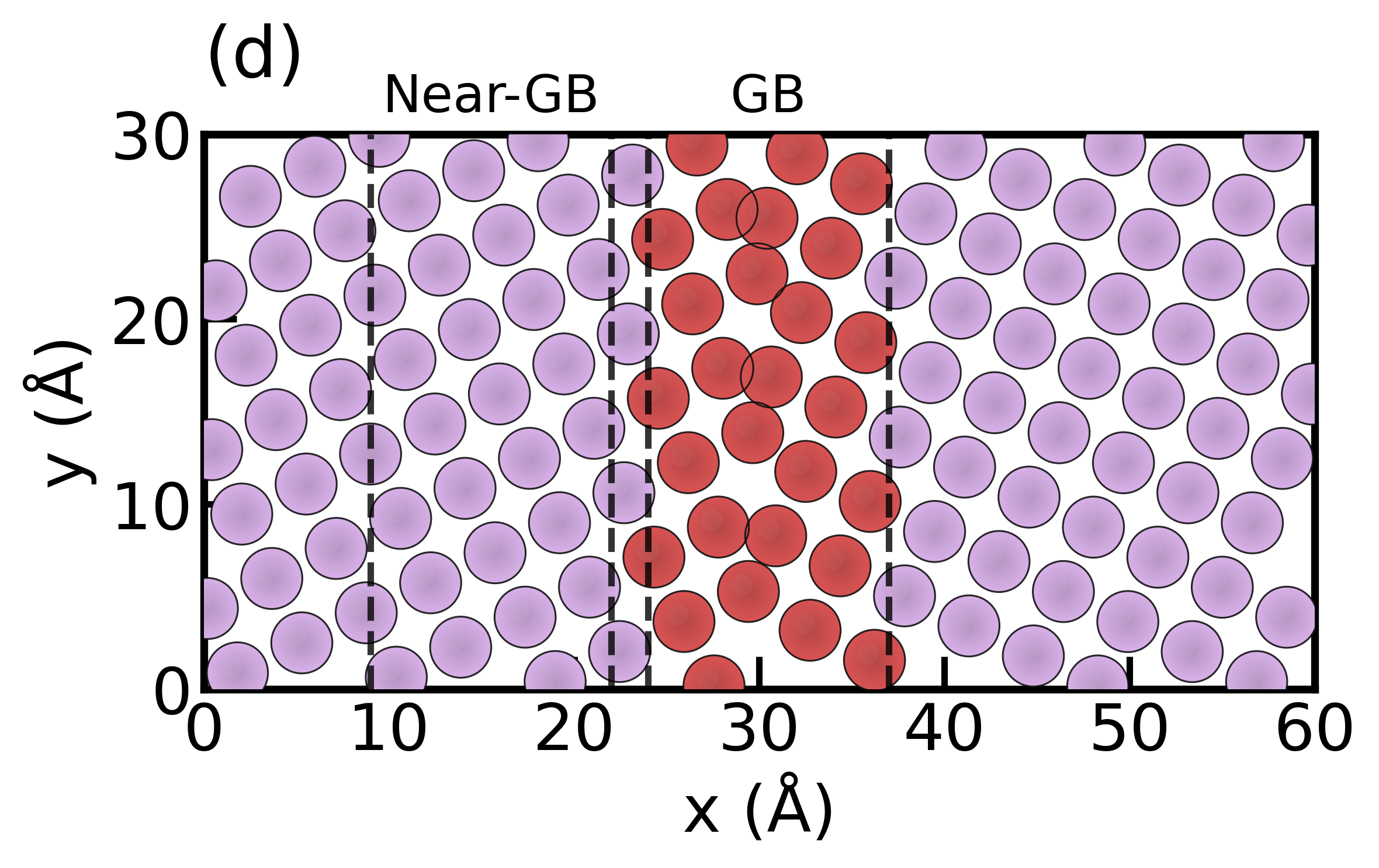}
\end{subfigure}

\caption{
Atomistic structures of symmetric tilt grain boundaries in silicon for misorientation angles of 
(a) $10^\circ$, (b) $20^\circ$, (c) $30^\circ$, and (d) $40^\circ$. The displayed structures correspond to the initial bicrystal configurations used as input for the molecular dynamics simulations.
}

\label{fig:MD_schematic}
\end{figure}

Grain boundaries play a dominant role in limiting thermal transport in nanocrystalline silicon, acting as strong phonon scattering centers that suppress long-wavelength heat-carrying modes and introduce interfacial thermal resistance~\cite{Wang2011NanoLett,Ju2013JAP,Li2019FrontPhys,Isotta2024AdvFunctMater}. To quantify how crystallographic disorder at such interfaces modifies phonon propagation, we construct atomistic models of symmetric tilt grain boundaries using a bicrystal geometry. The structures were built in a simulation box of dimensions $60 \times 30 \times 30~\text{\AA}^3$, containing 3189 atoms. Two crystalline grains were generated by assigning equal and opposite crystallographic rotations about the $z$ axis. Four misorientation angles were considered, corresponding to total tilt angles of $10^\circ$, $20^\circ$, $30^\circ$, and $40^\circ$, obtained by applying rotations of $\pm 5^\circ$, $\pm 10^\circ$, $\pm 15^\circ$, and $\pm 20^\circ$ to the two grains, respectively. This construction yields symmetric tilt grain boundaries normal to the $x$ direction, representing structurally disordered interfacial regions. By systematically varying the misorientation angle, we introduce a controlled increase in crystallographic disorder, enabling the investigation of how structural mismatch influences phonon scattering within a well-defined bicrystal geometry. This section focuses on crystallographic misorientation as a source of intrinsic interface disorder. More pronounced geometric roughness and highly disordered interface morphologies are examined separately in Section~\ref{sec:nemd} using non-equilibrium molecular dynamics simulations, where morphology-induced and non-equilibrium phonon scattering effects are explicitly analyzed. For spatially resolved calculations, the system was partitioned along the transport direction into a near grain-boundary region and a grain-boundary region, as illustrated in Figure~\ref{fig:MD_schematic}. The atomistic configurations were generated using Atomsk and visualized by directly reading the corresponding LAMMPS data files.

Atomistic simulations of both near-GB and GB  regions were performed using the GAP model (\texttt{Si\_Phonon.xml}) developed in this work from the bulk silicon dataset described in Sec.~\ref{sec:bulk}. Each bicrystal configuration was first fully relaxed via conjugate-gradient energy minimization, followed by periodic momentum zeroing to eliminate residual center-of-mass drift. For spatially resolved analysis, the simulation cell was partitioned along the transport direction into a near-GB region ($9 \le x \le 22$~\text{\AA}, $\sim$560 atoms) and a GB core region centered at the interface ($24 \le x \le 37$~\text{\AA}, $\sim$550 atoms). Finite-temperature equilibration was carried out using a staged protocol to stabilize the interfacial structure. Atomic velocities were initialized from a Maxwell--Boltzmann distribution at 100~K, followed by an NVT temperature ramp from 100~K to 300~K over 20~ps. An annealing step was then performed by equilibrating the system at 600~K for 20~ps to facilitate local atomic rearrangements within the grain-boundary region, after which the structure was cooled back to 300~K over an additional 20~ps. A final 40~ps NVE simulation at 300~K was performed to generate production trajectories free from thermostat-induced artifacts. Atomic positions, forces, and per-atom energies were recorded separately within the near-GB and interfacial regions to construct datasets representative of their distinct local environments. These data were used to train a MACE-based interatomic potential specifically for grain-boundary configurations, which was subsequently employed to calculate the phonon properties of the grain-boundary structures. The use of MACE for the lattice-dynamical calculations was motivated by its straightforward integration with the Phonopy and Phono3py workflows adopted in this work. Further details regarding dataset generation and MACE training are provided in the Supplementary Material (see Sec.~S2).

\begin{figure}[t]
    \centering
    \begin{subfigure}{0.8\linewidth}
        \centering
        \includegraphics[width=\linewidth]{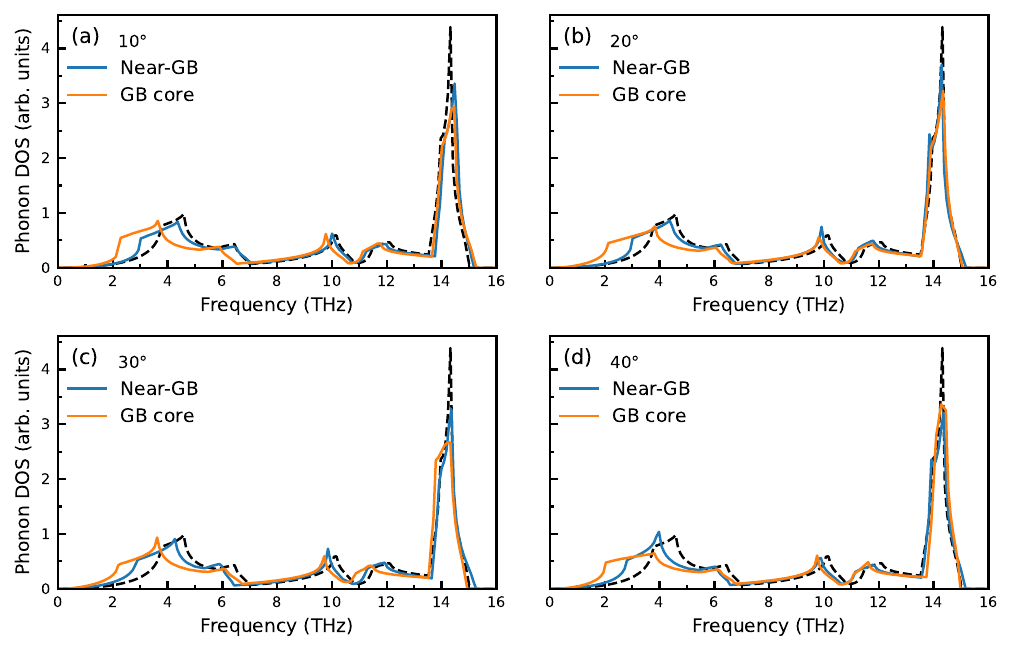}
        \label{fig:dos_gb_a}
    \end{subfigure}
    \caption{Phonon density of states of near-GB and grain-boundary core regions for different misorientation angles: (a) $10^\circ$, (b) $20^\circ$, (c) $30^\circ$, and (d) $40^\circ$. The dashed black curve denotes pristine bulk silicon}.
    \label{fig:dos_gb}
\end{figure}

Figure~\ref{fig:dos_gb}(a)--(d) compare the phonon DOS of the near-GB region and the GB region for misorientation angles of $10^\circ$, $20^\circ$, $30^\circ$, and $40^\circ$, respectively. In all cases, the near-GB region exhibits a DOS that is closer to pristine crystalline silicon than that of the GB core region, indicating a reduced influence of interfacial disorder away from the grain-boundary center. In contrast, the GB region shows stronger spectral modifications, particularly in the low- and mid-frequency ranges, manifested as peak broadening and redistribution of spectral weight. These deviations become increasingly pronounced as the misorientation angle increases, consistent with the larger structural disorder at the interface. The low-frequency acoustic modes, which dominate heat transport in bulk silicon, exhibit the strongest changes, consistent with reduced phonon coherence and partial spatial confinement near the boundary. High-frequency optical modes around $\sim 15$~THz remain present but display noticeable broadening and reduced intensity, reflecting stronger local scattering. Overall, the DOS evolution demonstrates that grain boundaries significantly reshape the vibrational landscape by perturbing long-wavelength acoustic modes and introducing disorder-related spectral features, providing a microscopic basis for the enhanced phonon scattering and reduced thermal conductivity in nanocrystalline silicon.

\begin{figure}[t]
    \centering
    \includegraphics[width=\linewidth]{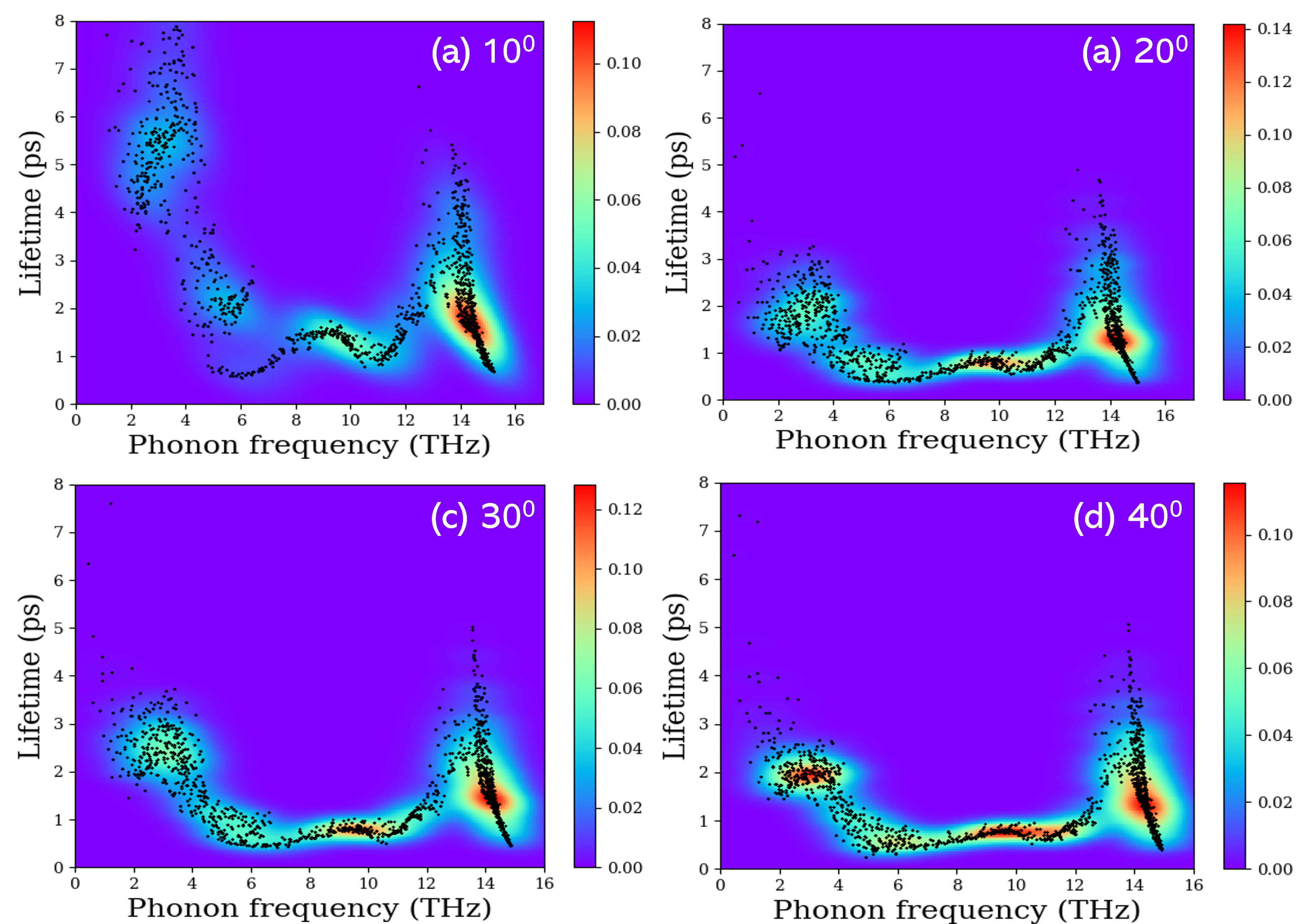}
    \caption{Phonon lifetimes of the grain-boundary core for different misorientation angles: (a) $10^\circ$, (b) $20^\circ$, (c) $30^\circ$, and (d) $40^\circ$ at 300~K.}
    \label{fig:gb_lifetime}
\end{figure}

Figure~\ref{fig:gb_lifetime}(a)--(d) shows the phonon lifetimes within the GB region for misorientation angles of $10^\circ$, $20^\circ$, $30^\circ$, and $40^\circ$ at 300~K, respectively. In each case, the lifetimes are substantially reduced relative to bulk silicon and remain on the order of a few picoseconds across a broad frequency range. Although the $10^\circ$ boundary exhibits slightly longer values, the overall similarity across angles indicates that once the interfacial region becomes structurally disordered, scattering is governed primarily by local atomic disorder rather than the precise misorientation. The persistently short lifetimes imply a significant reduction in phonon coherence length, particularly for long-wavelength acoustic modes, aligning with strong boundary-induced scattering and spatial confinement of heat-carrying phonons near the GB. For comparison, the phonon lifetimes in the near-GB regions are reported in the Supplementary Material (see Sec. S2). Unlike the GB core, they show minimal dependence on misorientation angle, indicating that the strong scattering observed at the interface is localized within the grain-boundary region.
\begin{figure}[t]
    \centering

    \begin{subfigure}{0.48\linewidth}
        \centering
        \includegraphics[width=\linewidth]{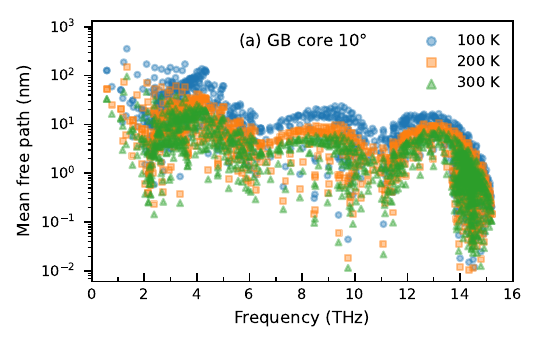}
        \label{fig:gb_mfp_freq_10}
    \end{subfigure}
    \hfill
    \begin{subfigure}{0.48\linewidth}
        \centering
        \includegraphics[width=\linewidth]{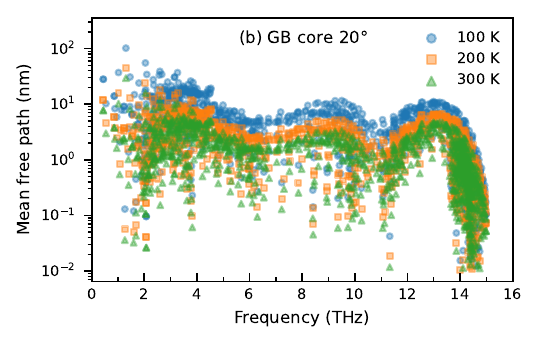}
        \label{fig:gb_mfp_freq_20}
    \end{subfigure}

    \caption{
    Phonon mean free path as a function of frequency within the grain-boundary core of nanocrystalline silicon for misorientation angles of (a) $10^\circ$ and (b) $20^\circ$, at different temperatures.
    }

    \label{fig:gb_mfp_freq}
\end{figure}

Figure~\ref{fig:gb_mfp_freq}(a) and (b) show the phonon mean free path as a function of frequency within the GB core for misorientation angles of $10^\circ$ and $20^\circ$, respectively. In both cases, the mean free paths are strongly suppressed across the full frequency range, indicating intense phonon scattering within the interfacial region. Low-frequency acoustic modes retain the largest values; however, even these are limited to a few tens of nanometers, substantially smaller than in bulk crystalline silicon. With increasing frequency, the mean free path decreases rapidly, reflecting enhanced scattering of mid- and high-frequency modes by structural disorder at the boundary. The $10^\circ$ grain boundary systematically exhibits slightly larger mean free paths, particularly in the low-frequency acoustic region, consistent with its comparatively more ordered interfacial structure and correspondingly weaker phonon scattering. In contrast, the $20^\circ$ grain boundary shows a stronger suppression of phonon propagation. The higher-angle grain boundaries ($30^\circ$ and $40^\circ$) exhibit qualitatively similar behavior to the $20^\circ$ case, indicating weak sensitivity to further increases in misorientation angle. The observed reduction is likely influenced by finite-size effects, local structural disorder, and interfacial phonon scattering, all of which can contribute to suppressing phonon transport relative to pristine bulk silicon. The temperature dependence further reveals a systematic reduction of mean free paths at elevated temperatures, consistent with enhanced anharmonic phonon--phonon scattering and reduced phonon coherence.


\begin{figure}[t]
    \centering

    \begin{subfigure}{0.48\linewidth}
        \centering
        \includegraphics[width=\linewidth]{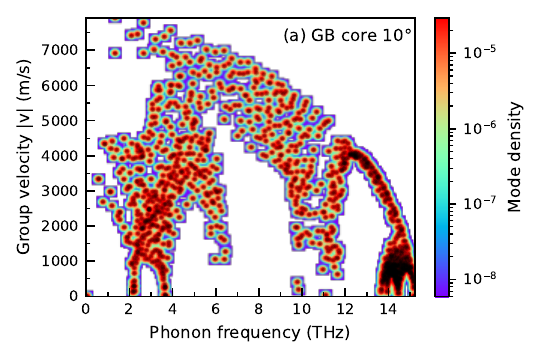}
        \label{fig:gb_vmap_10}
    \end{subfigure}
    \hfill
    \begin{subfigure}{0.48\linewidth}
        \centering
        \includegraphics[width=\linewidth]{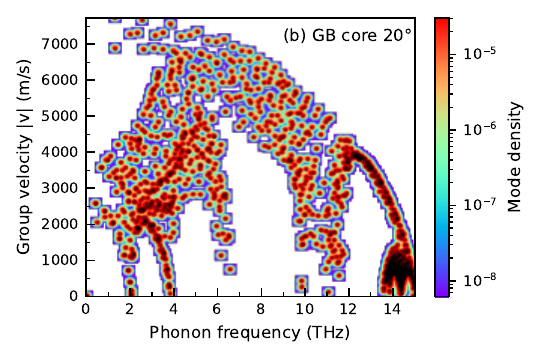}
        \label{fig:gb_vmap_20}
    \end{subfigure}

    \caption{Phonon group velocity magnitude as a function of phonon frequency within the grain-boundary core at 300~K for misorientation angles of (a) $10^\circ$ and (b) $20^\circ$.}
    
    \label{fig:gb_velocity_map}
\end{figure}

\begin{figure}[t]
    \centering
    \begin{subfigure}{0.48\linewidth}
        \centering
        \includegraphics[width=\linewidth]{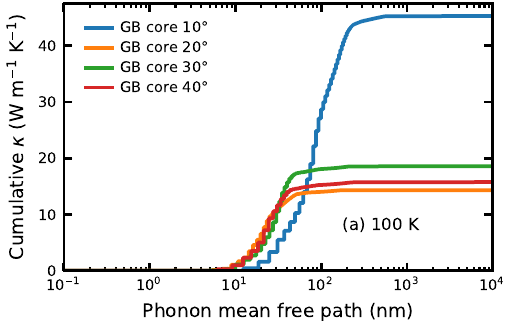}
        \label{fig:gb_kaccum_100K}
    \end{subfigure}
    \hfill
    \begin{subfigure}{0.48\linewidth}
        \centering
        \includegraphics[width=\linewidth]{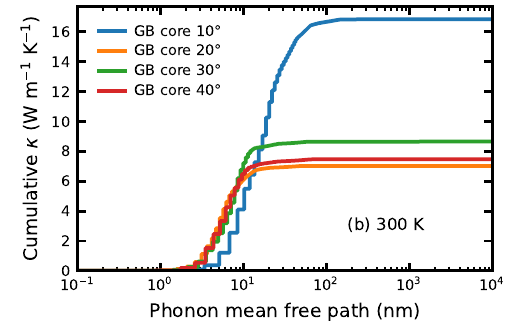}
        \label{fig:gb_kaccum_300K}
    \end{subfigure}

    \vspace{0.6em}

    \begin{subfigure}{0.48\linewidth}
        \centering
        \includegraphics[width=\linewidth]{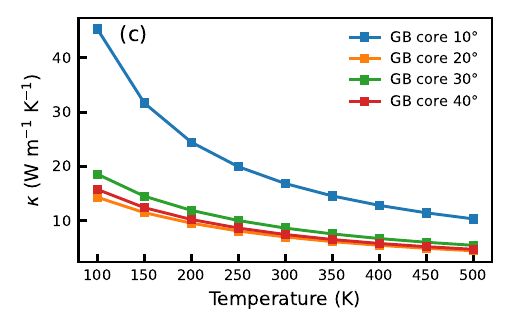}
        \label{fig:gb_kappa}
    \end{subfigure}
    \hfill
    \begin{subfigure}{0.48\linewidth}
        \centering
        \includegraphics[width=\linewidth]{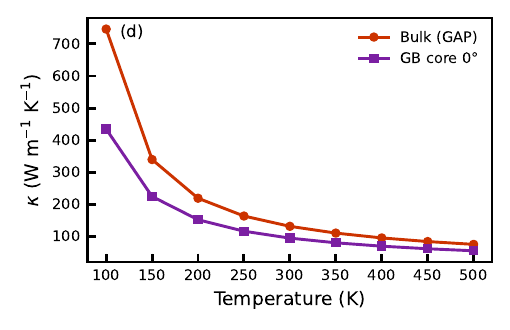}
        \label{fig:bulk_vs_0deg}
    \end{subfigure}

    \caption{Grain-boundary core transport characteristics in nanocrystalline silicon.
     Cumulative contribution to thermal transport as a function of phonon mean free path at (a) 100~K and
    (b) 300~K. (c) Temperature dependence of the thermal conductivity of the GB core for misorientation angles of $10^\circ$, $20^\circ$, $30^\circ$, and $40^\circ$.
    (d) Comparison of thermal conductivity between bulk pristine and the $0^\circ$ GB configuration.}
    
    \label{fig:gb_transport}
\end{figure}

Figure~\ref{fig:gb_velocity_map}(a) and (b) present the phonon group-velocity distributions within the GB core at 300~K for increasing misorientation angles. Compared with bulk silicon, the most noticeable changes occur in the acoustic frequency range, where distinct velocity branches become broader and less clearly separated. In particular, the low- and high-velocity branches between approximately 2 and 6~THz tend to merge, while the higher-velocity branch between 6 and 10~THz exhibits increased broadening. These features indicate a redistribution of phonon velocities within the grain-boundary region and are consistent with enhanced mode mixing associated with the local structural disorder.

Figure~\ref{fig:gb_transport}(a) and (b) show the cumulative thermal-conductivity contribution as a function of phonon mean free path within the GB region at 100~K and 300~K, respectively. In both cases, the cumulative curves saturate at relatively short mean free paths, indicating that long-propagation phonons contribute only weakly to heat transport inside the interface. Saturation occurs at slightly larger mean free paths at 100~K than at 300~K, consistent with reduced anharmonic phonon–phonon interactions and correspondingly longer coherence lengths at lower temperatures. In contrast, at 300~K, enhanced scattering shortens the effective propagation length and accelerates saturation. The weak variation with misorientation angle indicates that transport within the GB core is governed primarily by the local atomic configuration rather than the specific crystallographic tilt, although the $10^\circ$ boundary exhibits distinct behavior associated with its comparatively more ordered structure.
Figure~\ref{fig:gb_transport}(c) presents the temperature dependence of the GB thermal conductivity for misorientation angles of $10^\circ$, $20^\circ$, $30^\circ$, and $40^\circ$. In all cases, the conductivity remains far below that of bulk silicon and varies only weakly with temperature over the investigated range. The $10^\circ$ boundary exhibits slightly higher values than the larger-angle boundaries, consistent with its comparatively more coherent atomic arrangement. The low magnitude and weak temperature sensitivity are consistent with the limited propagation lengths observed in the interfacial region, indicating that heat transfer across the GB is constrained by its non-periodic atomic structure and the associated phonon–boundary scattering. Figure~\ref{fig:gb_transport}(d) compares the thermal conductivity of bulk silicon with a $0^\circ$ configuration, which serves as a reference system without a true grain boundary, but a slightly disordered interface due to the different thermal preparation treatment relative to crystalline domains, as described above for the other orientations. The results show that the $0^\circ$ case consistently exhibits lower thermal conductivity than bulk silicon across the entire temperature range. This reduction suggests that, despite the absence of misorientation, the presence of the interfacial region and finite-size effects in the bicrystal model contribute to additional phonon scattering. Nevertheless, the overall temperature dependence closely follows that of the bulk, confirming that the $0^\circ$ configuration behaves as a bulk-like reference, with larger thermal conductivity than other GB models.

\section{Machine-Learning Non-Equilibrium Molecular Dynamics}
\label{sec:nemd}

Interfacial heat transport across symmetric tilt grain boundaries is investigated using a machine-learning-based non-equilibrium molecular dynamics (ML-NEMD) approach. In this framework, a finite temperature gradient is imposed across the system to drive heat flow, and the resulting steady-state heat flux is computed. This method naturally captures full anharmonic phonon--phonon interactions as well as non-equilibrium effects, providing a comprehensive description of thermal transport at the interface. The thermal boundary resistance is extracted from the temperature discontinuity at the grain boundary, as obtained from the steady-state temperature profile.

\begin{figure}[t]
    \centering

    \begin{subfigure}{0.65\linewidth}
        \centering
        \includegraphics[width=\linewidth]{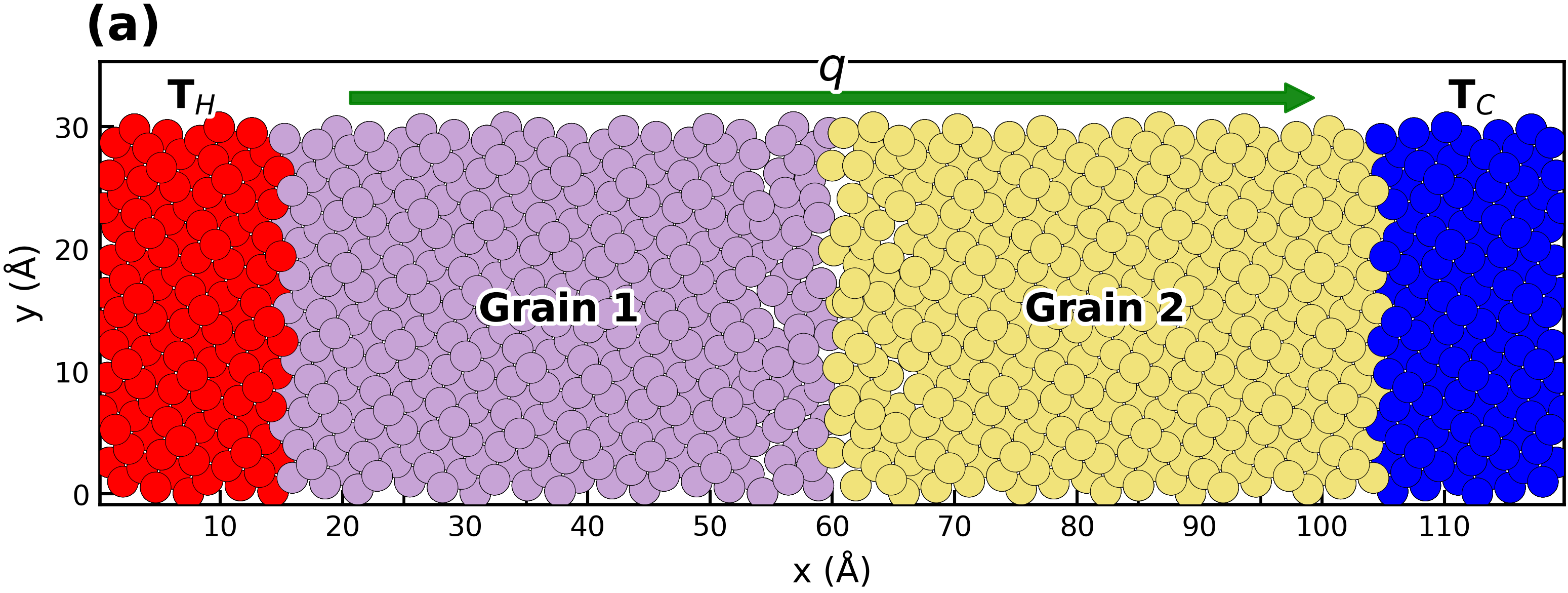}
    \end{subfigure}
    \hfill
    \begin{subfigure}{0.65\linewidth}
        \centering
        \includegraphics[width=\linewidth]{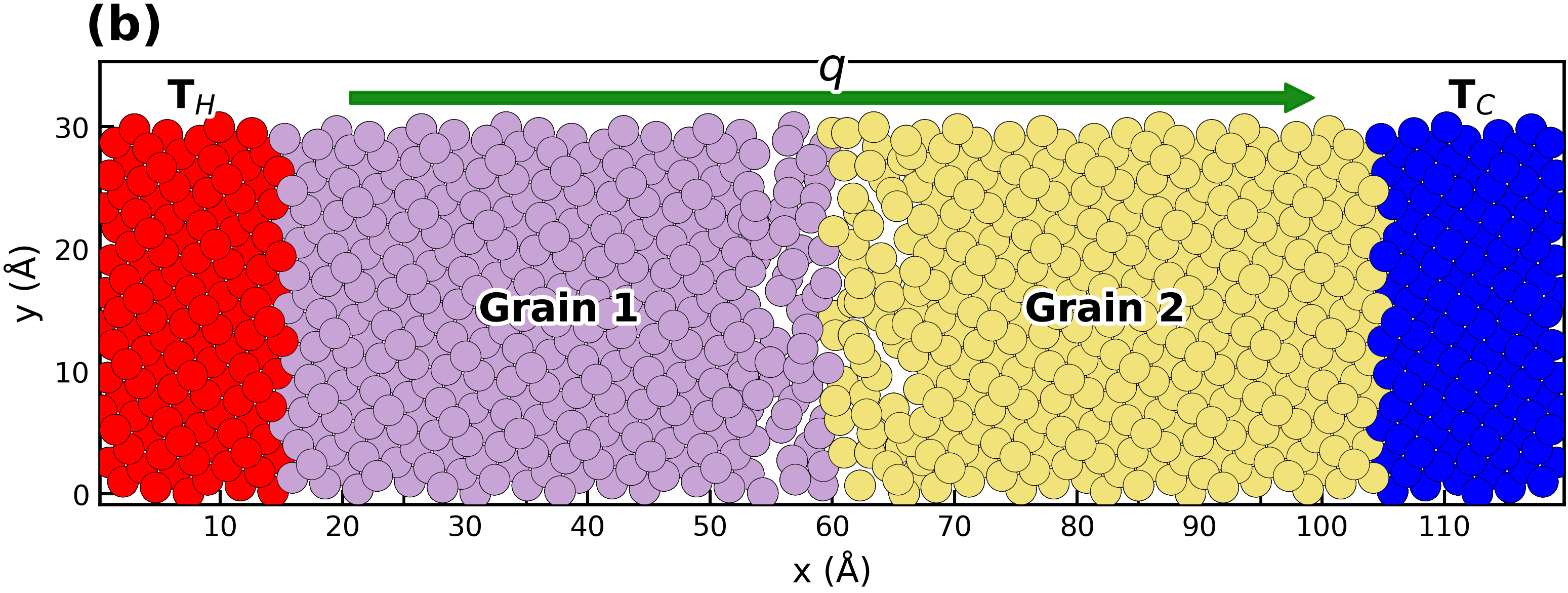}
    \end{subfigure}
    \begin{subfigure}{0.65\linewidth}
        \centering
        \includegraphics[width=\linewidth]{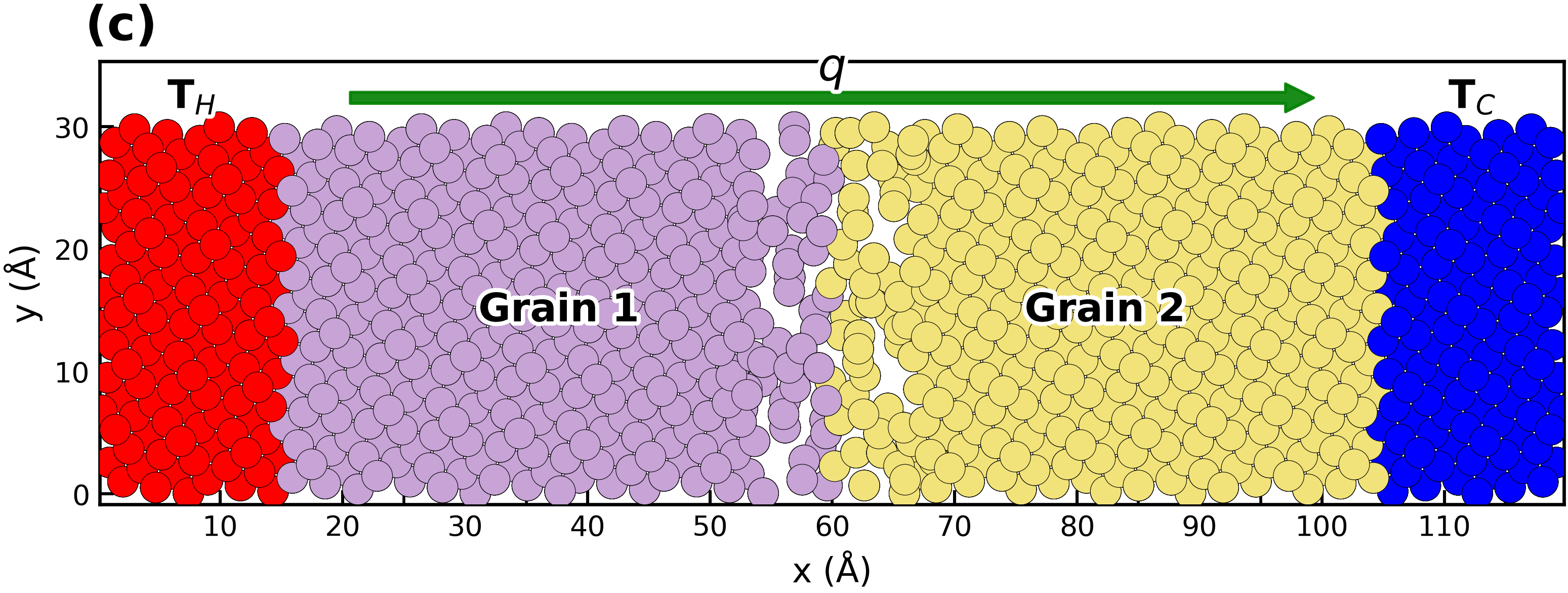}
    \end{subfigure}

\caption{Schematic of the non-equilibrium molecular dynamics setup used to compute interfacial thermal transport across a symmetric $20^\circ$ silicon grain boundary. 
(a) $A = 1$~\AA, (b) $A = 2$~\AA, and (c) $A = 3$~\AA\ correspond to different sinusoidal roughness amplitudes.}
    
    \label{fig:nemd_schematic}
\end{figure}

To directly quantify interfacial thermal transport across grain boundaries, ML-NEMD simulations were performed in LAMMPS using the QUIP/QUIPY interface. Unless otherwise stated, the simulations employed the GAP trained on the bulk silicon dataset (\texttt{Si\_Phonon.xml}). For comparison, selected ML-NEMD calculations were also carried out using classical SW and Tersoff potentials. Nanocrystalline silicon bicrystals containing symmetric tilt grain boundaries were constructed in a simulation cell of dimensions $120 \times 30 \times 30~\text{\AA}^3$, comprising 5350 atoms with periodic boundary conditions applied in all directions. The grain-boundary plane was positioned near the center of the cell ($x \approx 60$~\text{\AA}) and oriented normal to the heat-transport direction. The ML-NEMD simulations were designed to investigate strongly perturbed and roughened interfaces in order to capture realistic non-equilibrium scattering processes. In this context, controlled roughness was introduced locally within a grain-boundary slab of total thickness 12~\text{\AA} by imposing a sinusoidal atomic displacement field along the transport direction. From the generated \texttt{.lmp} data files, three roughness amplitudes, $A = 1$~\text{\AA}, $A = 2$~\text{\AA}, and $A = 3$~\text{\AA}, were applied, as illustrated in Figures.~\ref{fig:nemd_schematic}(a), (b), and (c), respectively. This procedure modifies only the local interfacial geometry while preserving crystalline order in the adjoining bulk regions. Starting from the rough bicrystal configuration, the system was first energy-minimized using a conjugate-gradient algorithm and subsequently equilibrated using a staged NVT protocol consisting of a 100~K hold for 20~ps, a ramp from 100~K to 300~K over 20~ps, a ramp from 300~K to 350~K over 20~ps, and a final 350~K hold for 40~ps to stabilize the grain-boundary structure. The equilibrated configuration was written to a restart file and used as the initial state for the NEMD production runs.

In the NEMD setup, two Langevin heat reservoirs of thickness 15~\text{\AA} were imposed at both ends of the simulation cell along the transport direction, maintaining temperatures of $T_\mathrm{hot}=400$~K and $T_\mathrm{cold}=300$~K on the left and right sides, respectively, while the remaining atoms evolved under microcanonical (NVE) dynamics. The steady-state heat flux was computed from the cumulative energy exchange between the system and the thermostats using the Langevin tally method. Spatially resolved temperature profiles were obtained by dividing the mobile region into bins along the transport direction and time-averaging the local kinetic temperature within each bin. Following an initial transient period, steady-state conditions were maintained for an additional 2~ns. The temperature drop across the grain boundary and the corresponding thermal boundary resistance were then determined from the resulting temperature profile. In all cases, the energy increases approximately linearly with time due to continuous energy exchange with the Langevin thermostats at the boundaries. The detailed NEMD workflow, together with additional results for roughness amplitudes of $A = 1$~\text{\AA} and $A =2$~\text{\AA}, is provided in the Supplementary Material (see Sec.~S3).

\begin{figure}[t]
    \centering
    \begin{subfigure}{0.48\linewidth}
        \centering
        \includegraphics[width=\linewidth]{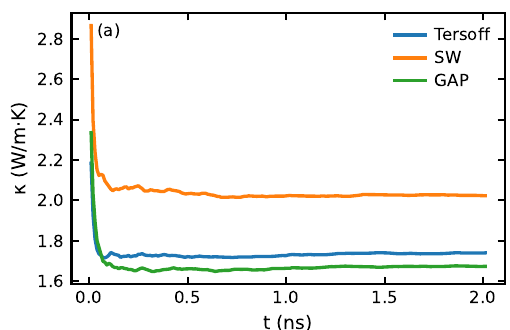}
        \label{fig:nemd_kappa}
    \end{subfigure}
    \hfill
    \begin{subfigure}{0.48\linewidth}
        \centering
        \includegraphics[width=\linewidth]{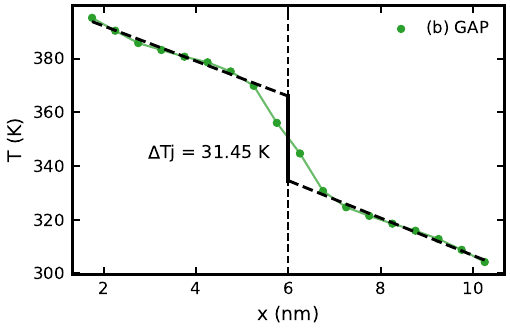}
        \label{fig:Tprofile_gap}
    \end{subfigure}

    \vspace{0.6em}

    \begin{subfigure}{0.48\linewidth}
        \centering
        \includegraphics[width=\linewidth]{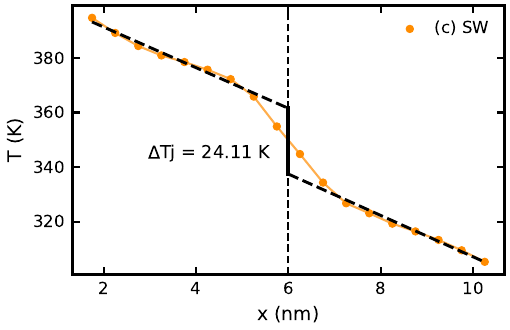}
        \label{fig:Tprofile_sw}
    \end{subfigure}
    \hfill
    \begin{subfigure}{0.48\linewidth}
        \centering
        \includegraphics[width=\linewidth]{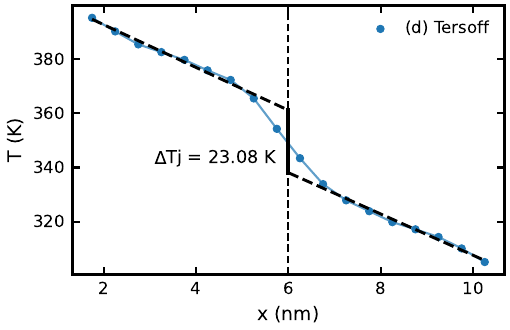}
        \label{fig:Tprofile_ter}
    \end{subfigure}
\caption{Non-equilibrium molecular dynamics simulations for $A = 3$~\text{\AA}. 
(a) Time evolution of the effective thermal conductivity $\kappa$ obtained from the NEMD simulations using the Tersoff, Stillinger--Weber, and GAP interatomic potentials. 
(b--d) Steady-state temperature profiles along the transport direction.}

    \label{fig:nemd_results_A3}
\end{figure}
Figures~\ref{fig:nemd_results_A3}(b)--(d) present the steady-state temperature profiles obtained using the GAP, Stillinger--Weber, and Tersoff interatomic potentials, respectively. In all cases, linear temperature gradients develop in the bulk regions on either side of the grain boundary, while a clear temperature discontinuity is observed at the interface. The interfacial temperature jump, $\Delta T_j$, is determined by linearly fitting the temperature profiles in the bulk regions and extrapolating these fits to the grain-boundary position $x_{\mathrm{GB}}$, according to
\begin{equation}
\Delta T_j = T_{\mathrm{R}}(x_{\mathrm{GB}}) - T_{\mathrm{L}}(x_{\mathrm{GB}}),
\end{equation}
where $T_{\mathrm{L}}$ and $T_{\mathrm{R}}$ denote the extrapolated temperatures on the left and right sides of the grain boundary, respectively. The magnitude of $\Delta T_j$ depends on the choice of interatomic potential and reflects differences in the predicted phonon scattering strength at the grain boundary. The thermal boundary resistance (TBR) is obtained from the ratio of the interfacial temperature jump to the steady-state heat flux, according to
\begin{equation}
\mathrm{TBR} = \frac{|\Delta T_j|}{|J_x|}.
\end{equation}
In a steady state, the heat flux $J_x$ and the temperature gradient are related through Fourier’s law, $J_x = -\kappa \nabla T$, which allows an effective thermal conductivity to be extracted under identical nonequilibrium driving conditions.
Figure~\ref{fig:nemd_results_A3}(a) shows the evolution of $\kappa(t)$ for $A=3$~\text{\AA}. The higher steady-state values obtained with the SW and Tersoff potentials indicate weaker effective phonon scattering relative to GAP. Empirical bond-order models typically provide a simplified description of anharmonic phonon–phonon interactions, which result in longer phonon lifetimes and consequently higher thermal conductivities, as discussed above. In contrast, the machine-learned GAP model offers a more flexible representation of the potential energy surface, leading to enhanced anharmonic scattering and lower effective conductivity. These differences in lattice transport directly affect the steady-state heat flux entering the definition of thermal boundary resistance. Figures~\ref{fig:nemd_results_A3}(b)--(d) show the corresponding steady-state temperature profiles obtained using the GAP, SW, and Tersoff potentials, respectively. In all cases, linear temperature gradients develop in the bulk regions on either side of the grain boundary, while a clear temperature discontinuity appears at the interface. The magnitude of the temperature jump differs among the potentials: the GAP model predicts the largest discontinuity ($\Delta T_j \approx 31$~K), whereas smaller jumps are obtained with the SW ($\Delta T_j \approx 24$~K) and Tersoff ($\Delta T_j \approx 23$~K) potentials. This behavior indicates stronger resistance to vibrational energy transfer across the interface in the GAP description, consistent with its lower steady-state heat flux and enhanced anharmonic scattering.
\begin{table}[htbp]
\centering
\setlength{\tabcolsep}{4pt}
\footnotesize
\caption{Steady-state heat flux $J_x$, temperature jump $\Delta T_j$, effective thermal conductivity $\kappa$, and TBR for different roughness amplitudes $A$ and interatomic potentials. Experimental TBR ranges are included for qualitative comparison.}
\label{tab:nemd_summary}
\begin{tabular}{lccccc}
\hline
Potential & $A$ (\AA) & $|J_x|$ ($10^{10}$ W m$^{-2}$) & $\Delta T_j$ (K) & $\kappa$ (W m$^{-1}$K$^{-1}$) & TBR (m$^2$ K GW$^{-1}$) \\
\hline

\multirow{3}{*}{GAP}
    & 1 & 1.97 &  9.77 & 1.82 & 0.50 \\
    & 2 & 1.85 & 12.68 & 1.72 & 0.68 \\
    & 3 & 1.89 & 31.45 & 1.67 & 1.66 \\
\hline

\multirow{3}{*}{SW}
    & 1 & 2.74 & 14.18 & 2.69 & 0.52 \\
    & 2 & 2.63 & 19.23 & 2.45 & 0.73 \\
    & 3 & 2.23 & 24.11 & 2.02 & 1.08 \\
\hline

\multirow{3}{*}{Tersoff}
    & 1 & 2.45 & 10.78 & 2.39 & 0.44 \\
    & 2 & 2.32 & 16.24 & 2.13 & 0.70 \\
    & 3 & 1.97 & 23.08 & 1.74 & 1.17 \\
\hline

\multicolumn{6}{c}{\textit{Experimental TBR (m$^2$ K GW$^{-1}$)}} \\
\hline
\multicolumn{6}{c}{$\Sigma3$: 0.3--0.5 \qquad Rough GB: 0.5--1 \qquad Nano-twins: 1.7--2.3} \\
\hline

\end{tabular}
\end{table}
The values reported in Table~\ref{tab:nemd_summary} reveal a clear dependence of interfacial thermal transport on both the grain-boundary roughness amplitude and the underlying interatomic potential. For all three potentials, increasing the roughness amplitude systematically increases the temperature discontinuity and TBR, indicating progressively stronger phonon scattering at increasingly disordered interfaces. This trend is particularly pronounced for the GAP model, where the TBR increases from 0.50 to 1.66~m$^2$ K GW$^{-1}$ as the interface roughness increases from 1 to 3~\text{\AA}. In comparison, the SW and Tersoff potentials exhibit a more moderate increase, reaching 1.08 and 1.17~m$^2$ K GW$^{-1}$, respectively, at $A=3$~\text{\AA}. 

For the smaller roughness amplitudes ($A=1$ and $A=2$~\text{\AA}), the three models predict TBR values of comparable magnitude. However, this apparent agreement does not imply an equivalent description of interfacial heat transport, since the corresponding heat fluxes ($J_x$) and temperature discontinuities ($\Delta T_j$) differ noticeably among the potentials. The similarity in TBR, therefore, arises from a compensation between $\Delta T_j$ and $J_x$, rather than from an identical representation of the underlying phonon-scattering mechanisms. More pronounced differences emerge for the largest roughness amplitude ($A=3$~\text{\AA}), where the GAP model predicts a substantially larger temperature jump and correspondingly higher TBR than the SW and Tersoff potentials. These results indicate that the choice of interatomic potential has a significant impact on the predicted interfacial thermal transport, particularly for highly disordered grain-boundary morphologies. The effective thermal conductivity $\kappa$ shows moderate variations with roughness amplitude for all potentials, with smaller variations observed for GAP ($\sim$0.15~W\,m$^{-1}$\,K$^{-1}$) compared to SW and Tersoff ($\sim$0.6--0.7~W\,m$^{-1}$\,K$^{-1}$). This behavior reflects the increasing influence of grain-boundary scattering on heat transport across the finite simulation cell, rather than a modification of the intrinsic thermal conductivity of bulk silicon. The SW and Tersoff potentials exhibit increasingly different responses as the roughness amplitude increases, reflecting the sensitivity of the predicted interfacial transport properties to the choice of interatomic potential. These trends are further supported by recent experimental measurements of thermal transport at silicon grain boundaries, which report thermal boundary resistance values ranging from approximately 0.3--2.3~m$^2$ K GW$^{-1}$, with typical values around $\sim$0.5~m$^2$ K GW$^{-1}$ for smooth grain boundaries and significantly higher values (up to $\sim$1.5~m$^2$ K GW$^{-1}$) in the presence of structural complexity such as nanotwinning or increased interface roughness~\cite{Isotta2024AdvFunctMater}. The increase in TBR with interfacial complexity observed experimentally is consistent with the higher TBR predicted by the GAP potential at larger roughness amplitudes, indicating that the machine-learning model captures enhanced phonon scattering associated with complex grain-boundary morphologies.

\section{Conclusions}

In summary, we developed a machine-learning–based multiscale framework for investigating phonon-mediated thermal transport in bulk and nanocrystalline silicon by integrating GAP and MACE interatomic potentials with lattice-dynamical calculations (Phonopy/Phono3py) and non-equilibrium molecular dynamics. By directly employing MLIP-derived forces to compute second- and third-order force constants, the approach provides a consistent and accurate description of phonon dispersions, lifetimes, and lattice thermal conductivity beyond conventional empirical potentials, thereby establishing a reliable bulk reference for interfacial transport analysis. Complementary NEMD simulations reveal a finite temperature discontinuity at grain boundaries and demonstrate that the resulting thermal boundary resistance is highly sensitive to the underlying interatomic potential and its description of phonon scattering. Overall, these results highlight that predictive modeling of interfacial heat transport requires a unified and physically consistent treatment of both harmonic and anharmonic lattice dynamics within atomistic frameworks capable of capturing structural complexity. The results further indicate that interfacial thermal resistance is strongly influenced by the interplay between grain-boundary structure and the underlying interatomic potential, highlighting the importance of accurately describing phonon scattering in low dimensional systems.

\section*{Acknowledgements}
The authors acknowledge support from the EU Horizon Europe MAGNIFIC project, which has received funding from the European Union's Horizon Europe research and innovation under Grant Agreement No.~101091968. The authors thank Dr.~Yangyu Guo (Harbin Institute of Technology) for valuable discussions.

\section*{Supplementary Material}
The Supplementary Material provides additional details on the computational workflows and validation procedures used in this work. It includes the bulk, grain-boundary, and NEMD simulation protocols, together with convergence tests and additional temperature profiles supporting the main results.

\section*{Data Availability}

The machine-learning interatomic potential and phonon transport data supporting this work are openly available in the GitHub repository:
\url{https://github.com/Hrezgui/MLIP-Silicon-Interfacial-Thermal-Transport}.

The repository includes the GAP potential used in this study (\texttt{GAP\_Si\_Phonon.xml}) and its associated sparse descriptor file, compatible with the QUIP code and with LAMMPS through the \texttt{pair\_style quip} interface. It also contains the second- and third-order interatomic force constants (\texttt{fc2.hdf5} and \texttt{fc3.hdf5}), harmonic force constants, and Phono3py thermal-transport data for the symmetric tilt grain boundaries with misorientation angles of $10^\circ$, $20^\circ$, $30^\circ$, and $40^\circ$.

\bibliographystyle{apsrev4-2}
\makeatletter
\def\bibsection{\section*{References}}
\makeatother
\bibliography{REF}

@article{Cahill2014APR,
  author  = {Cahill, David G. and Braun, Philip V. and Chen, Gang and Clarke, David R. and Fan, Shanhui and Goodson, Kenneth E. and Keblinski, Pawel and King, William P. and Mahan, Gerald D. and Majumdar, Arun and Maris, Humphrey J. and Phillpot, Simon R. and Pop, Eric and Shi, Li},
  title   = {Nanoscale thermal transport. II. 2003--2012},
  journal = {Applied Physics Reviews},
  year    = {2014},
  volume  = {1},
  pages   = {011305}
}

@article{Pop2010NanoRes,
  author  = {Pop, Eric},
  title   = {Energy dissipation and transport in nanoscale devices},
  journal = {Nano Research},
  year    = {2010},
  volume  = {3},
  pages   = {147--169}
}

@book{Chen2005Book,
  author    = {Chen, Gang},
  title     = {Nanoscale Energy Transport and Conversion},
  publisher = {Oxford University Press},
  year      = {2005}
}

@article{Minnich2011PRL,
  author  = {Minnich, Austin J. and Johnson, John A. and Schmidt, Aaron J. and Esfarjani, Keivan and Dresselhaus, Mildred S. and Nelson, Keith A. and Chen, Gang},
  title   = {Thermal conductivity spectroscopy technique to measure phonon mean free paths},
  journal = {Physical Review Letters},
  year    = {2011},
  volume  = {107},
  pages   = {095901}
}

@article{Broido2007APL,
  author  = {Broido, David A. and Malorny, Michael and Birner, Gernot and Mingo, Natalio and Stewart, David A.},
  title   = {Intrinsic lattice thermal conductivity of semiconductors from first principles},
  journal = {Applied Physics Letters},
  year    = {2007},
  volume  = {91},
  pages   = {231922}
}

@article{Esfarjani2011PRB,
  author  = {Esfarjani, Keivan and Chen, Gang and Stokes, Harold T.},
  title   = {Heat transport in silicon from first-principles calculations},
  journal = {Physical Review B},
  year    = {2011},
  volume  = {84},
  number  = {8},
  pages   = {085204},
  doi     = {10.1103/PhysRevB.84.085204},
}

@article{Mingo2003PRB,
  author  = {Mingo, Natalio},
  title   = {Calculation of nanowire thermal conductivity using complete phonon dispersion relations},
  journal = {Physical Review B},
  year    = {2003},
  volume  = {68},
  pages   = {113308}
}

@article{Watanabe2007JApplPhys,
  author  = {Watanabe, Taku and Ni, Boris and Phillpot, Simon R. and Schelling, Patrick K. and Keblinski, Pawel},
  title   = {Thermal conductance across grain boundaries in diamond from molecular dynamics simulation},
  journal = {Journal of Applied Physics},
  year    = {2007},
  volume  = {102},
  number  = {6},
  pages   = {063503},
  doi     = {10.1063/1.2779289},
}

@article{Gordiz2015NJP,
  author  = {Gordiz, Kaveh and Henry, Asegun},
  title   = {A formalism for calculating the modal contributions to thermal interface conductance},
  journal = {New Journal of Physics},
  year    = {2015},
  volume  = {17},
  pages   = {103002}
}

@article{Plimpton1995JCP,
  author  = {Plimpton, Steve},
  title   = {Fast Parallel Algorithms for Short-Range Molecular Dynamics},
  journal = {Journal of Computational Physics},
  year    = {1995},
  volume  = {117},
  pages   = {1--19}
}

@article{Volz1999PRB,
  author  = {Volz, Sebastian G. and Chen, Gang},
  title   = {Molecular-dynamics simulation of thermal conductivity of silicon crystals},
  journal = {Physical Review B},
  year    = {1999},
  volume  = {61},
  pages   = {2651--2656}
}

@article{Henry2009PRB,
  author  = {Henry, Asegun and Chen, Gang},
  title   = {Spectral phonon transport properties of silicon from molecular dynamics simulations},
  journal = {Physical Review B},
  year    = {2009},
  volume  = {79},
  pages   = {144305}
}

@article{Stillinger1985PRB,
  author  = {Stillinger, Frank H. and Weber, Thomas A.},
  title   = {Computer simulation of local order in condensed phases of silicon},
  journal = {Physical Review B},
  year    = {1985},
  volume  = {31},
  pages   = {5262--5271}
}

@article{Tersoff1988PRB,
  author  = {Tersoff, J.},
  title   = {New empirical approach for the structure and energy of covalent systems},
  journal = {Physical Review B},
  year    = {1988},
  volume  = {37},
  pages   = {6991--7000}
}

@article{Behler2007PRL,
  author  = {Behler, J{\"o}rg and Parrinello, Michele},
  title   = {Generalized Neural-Network Representation of High-Dimensional Potential-Energy Surfaces},
  journal = {Physical Review Letters},
  year    = {2007},
  volume  = {98},
  pages   = {146401}
}

@article{Bartok2010PRL,
  title = {Gaussian Approximation Potentials: The Accuracy of Quantum Mechanics, without the Electrons},
  author = {Bart\'ok, Albert P. and Payne, Mike C. and Kondor, Risi and Cs\'anyi, G\'abor},
  journal = {Phys. Rev. Lett.},
  volume = {104},
  issue = {13},
  pages = {136403},
  numpages = {4},
  year = {2010},
  month = {Apr},
  publisher = {American Physical Society},
  doi = {10.1103/PhysRevLett.104.136403},
  url = {https://link.aps.org/doi/10.1103/PhysRevLett.104.136403}
}

@article{Bartok2018PRX,
  author  = {Bart{\'o}k, Albert P. and De, Sandip and Poelking, Christian and Bernstein, Noam and Kermode, James R. and Cs{\'a}nyi, G{\'a}bor},
  title   = {Machine learning a general-purpose interatomic potential for silicon},
  journal = {Physical Review X},
  year    = {2018},
  volume  = {8},
  number  = {4},
  pages   = {041048},
  doi     = {10.1103/PhysRevX.8.041048},
  url     = {https://link.aps.org/doi/10.1103/PhysRevX.8.041048}
}

@article{Bartok2013PhysRevB,
  author  = {Bart{\'o}k, Albert P. and Kondor, Risi and Cs{\'a}nyi, G{\'a}bor},
  title   = {On representing chemical environments},
  journal = {Physical Review B},
  year    = {2013},
  volume  = {87},
  number  = {18},
  pages   = {184115},
  doi     = {10.1103/PhysRevB.87.184115},
}

@article{Kermode2020-wu,
  title    = "f90wrap: an automated tool for constructing deep Python
              interfaces to modern Fortran codes",
  author   = "Kermode, James R",
  journal  = "J. Phys. Condens. Matter",
  month    =  mar,
  year     =  2020,
  language = "en",
  issn     = "0953-8984, 1361-648X",
  pmid     = "32209737",
  doi      = "10.1088/1361-648X/ab82d2"
}

@article{Deringer2021ChemRev,
  author  = {Deringer, Volker L. and Bartók, Albert P. and Bernstein, Noam and Wilkins, David M. and Ceriotti, Michele and Csányi, Gábor},
  title   = {Gaussian Process Regression for Materials and Molecules},
  journal = {Chemical Reviews},
  year    = {2021},
  volume  = {121},
  number  = {16},
  pages   = {10073--10141},
  doi     = {10.1021/acs.chemrev.1c00022},
  publisher = {American Chemical Society}
}

@article{Caro2019PhysRevB,
  author  = {Caro, Miguel A.},
  title   = {Optimizing many-body atomic descriptors for enhanced computational performance of machine learning based interatomic potentials},
  journal = {Physical Review B},
  year    = {2019},
  volume  = {100},
  number  = {2},
  pages   = {024112},
  doi     = {10.1103/PhysRevB.100.024112},
}

@article{Batatia2022NeurIPS,
  author  = {Batatia, Ilyes and Kov{\'a}cs, D{\'a}vid P. and Simm, Gregor N. C. and Ortner, Christoph and Cs{\'a}nyi, G{\'a}bor},
  title   = {MACE: Higher Order Equivariant Message Passing Neural Networks for Fast and Accurate Force Fields},
  journal = {Advances in Neural Information Processing Systems},
  year    = {2022},
  volume  = {35},
  pages   = {11423--11436},
}

@article{Kovacs2023JCP,
  author  = {Kov{\'a}cs, D{\'a}vid P{\'e}ter and Batatia, Ilyes and Arany, Eszter S{\'a}ra and Cs{\'a}nyi, G{\'a}bor},
  title   = {Evaluation of the {MACE} force field architecture: From medicinal chemistry to materials science},
  journal = {The Journal of Chemical Physics},
  year    = {2023},
  volume  = {159},
  number  = {4},
  pages   = {044118},
  doi     = {10.1063/5.0155322},
}

@article{Guo2018PRB,
  author  = {Guo, Yangyu and Jou, David and Wang, Moran},
  title   = {Nonequilibrium thermodynamics of phonon hydrodynamic model for nanoscale heat transport},
  journal = {Physical Review B},
  year    = {2018},
  volume  = {98},
  number  = {10},
  pages   = {104304},
  doi     = {10.1103/PhysRevB.98.104304},
}

@article{PhysRevLett.110.265506,
  title = {Direct Solution to the Linearized Phonon Boltzmann Equation},
  author = {Chaput, Laurent},
  journal = {Phys. Rev. Lett.},
  volume = {110},
  issue = {26},
  pages = {265506},
  numpages = {5},
  year = {2013},
  month = {Jun},
  publisher = {American Physical Society},
  doi = {10.1103/PhysRevLett.110.265506},
  url = {https://link.aps.org/doi/10.1103/PhysRevLett.110.265506}
}

@article{kz9s-y611,
  title = {Scattering matrix formalism for interface heat conduction based on Monte Carlo evaluation of the phonon Boltzmann transport equation},
  author = {Li, Yifei and Chen, Yichong and Tang, Guihua and Gibelli, Livio and Borg, Matthew K.},
  journal = {Phys. Rev. B},
  volume = {113},
  issue = {20},
  pages = {205304},
  numpages = {14},
  year = {2026},
  month = {May},
  publisher = {American Physical Society},
  doi = {10.1103/kz9s-y611},
  url = {https://link.aps.org/doi/10.1103/kz9s-y611}
}

@article{Zhang2025APL,
  author  = {Zhang, Chuang and Wu, Lei},
  title   = {Theoretical studies of transient hydrodynamic phonon transport in two-dimensional disk geometry},
  journal = {Applied Physics Letters},
  year    = {2025},
  volume  = {126},
  number  = {3},
  pages   = {032201},
  doi     = {10.1063/5.0248153},
}

@article{Shan2025PRB,
  author  = {Shan, Shuyue and Zhang, Zhongwei and Lu, Shuang and Volz, Sebastian and Chen, Jie},
  title   = {Generation of interfacial phonon modes and their contribution to thermal transport across the {GaN}/{ZnO} interface},
  journal = {Physical Review B},
  year    = {2025},
  volume  = {112},
  number  = {15},
  pages   = {155302},
  doi     = {10.1103/PhysRevB.112.155302},
}

@article{Zhang2018PRLDeepMD,
  author  = {Zhang, Linfeng and Han, Jiequn and Wang, Han and Car, Roberto and E, Weinan},
  title   = {Deep Potential Molecular Dynamics: A Scalable Model with the Accuracy of Quantum Mechanics},
  journal = {Physical Review Letters},
  year    = {2018},
  volume  = {120},
  number  = {14},
  pages   = {143001},
  doi     = {10.1103/PhysRevLett.120.143001},
}

@article{Sledzinska2020AFM,
  author  = {Sledzinska, Marianna and Graczykowski, Bartlomiej and Maire, Jeremie and Chavez-Angel, Emigdio and Sotomayor-Torres, Clivia M. and Alzina, Francesc},
  title   = {2D Phononic Crystals: Progress and Prospects in Hypersound and Thermal Transport Engineering},
  journal = {Advanced Functional Materials},
  year    = {2020},
  volume  = {30},
  number  = {8},
  pages   = {1904434},
  doi     = {10.1002/adfm.201904434},
}

@article{Reig2022AdvMater,
  author  = {Saleta Reig, David and Varghese, Sebin and Farris, Roberta and Block, Alexander and Mehew, Jake D. and Hellman, Olle and Wo{\'z}niak, Pawe{\l} and Sledzinska, Marianna and El Sachat, Alexandros and Ch{\'a}vez-{\'A}ngel, Emigdio and Valenzuela, Sergio O. and van Hulst, Niek F. and Ordej{\'o}n, Pablo and Zanolli, Zeila and Sotomayor Torres, Clivia M. and Verstraete, Matthieu J. and Tielrooij, Klaas-Jan},
  title   = {Unraveling Heat Transport and Dissipation in Suspended {MoSe}$_2$ from Bulk to Monolayer},
  journal = {Advanced Materials},
  year    = {2022},
  volume  = {34},
  number  = {10},
  pages   = {2108352},
  doi     = {10.1002/adma.202108352}
}

@article{Rajabpour2025IJTS,
  author  = {Rajabpour, Ali and Mortazavi, Bohayra and Mirchi, Pedram and El Hajj, Julien and Guo, Yangyu and Zhuang, Xiaoying and Merabia, Samy},
  title   = {Accurate estimation of interfacial thermal conductance between silicon and diamond enabled by a machine learning interatomic potential},
  journal = {International Journal of Thermal Sciences},
  year    = {2025},
  volume  = {214},
  pages   = {109876},
  doi     = {10.1016/j.ijthermalsci.2025.109876},
}

@article{Chen2021NRP,
  author  = {Chen, Gang},
  title   = {Non-{Fourier} phonon heat conduction at the microscale and nanoscale},
  journal = {Nature Reviews Physics},
  year    = {2021},
  volume  = {3},
  number  = {8},
  pages   = {555--569},
  doi     = {10.1038/s42254-021-00334-1},
}

@article{Raciti2025AdvSci,
  author  = {Raciti, Grazia and Abad, Bego{\~n}a and Dettori, Riccardo and Sen, Raja and K. Sivan, Aswathi and Sojo-Gordillo, Jose M. and Vast, Nathalie and Rurali, Riccardo and Melis, Claudio and Sjakste, Jelena and Zardo, Ilaria},
  title   = {Unraveling Energy Flow Mechanisms in Semiconductors by Ultrafast Spectroscopy: Germanium as a Case Study},
  journal = {Advanced Science},
  year    = {2025},
  pages   = {e15470},
  doi     = {10.1002/advs.202515470},
}

@article{Albrigi2024APL,
  author  = {Albrigi, Tommaso and Rurali, Riccardo},
  title   = {Phonon transport across {GaAs}/{Ge} heterojunctions by nonequilibrium molecular dynamics},
  journal = {Applied Physics Letters},
  year    = {2024},
  volume  = {124},
  number  = {10},
  pages   = {102203},
  doi     = {10.1063/5.0191692},
}

@article{Farzadian2024APL,
  author  = {Farzadian, Omid and Sekerbayev, Kairolla and Wang, Yanwei and Utegulov, Zhandos N.},
  title   = {Nanoscale spatially resolved thermal transport in nanocrystalline {3C-SiC}},
  journal = {Applied Physics Letters},
  year    = {2024},
  volume  = {124},
  number  = {23},
  pages   = {232203},
  doi     = {10.1063/5.0206189},
}

@article{Guo2025JAP,
  author  = {Guo, Liben and Liu, Yuanbin and Yang, Lei and Cao, Bingyang},
  title   = {Lattice dynamics modeling of thermal transport in solids using machine-learned atomic cluster expansion potentials: A tutorial},
  journal = {Journal of Applied Physics},
  year    = {2025},
  volume  = {137},
  number  = {8},
  pages   = {081101},
  doi     = {10.1063/5.0251119},
  
}

@article{Khot2025JAP,
  author  = {Khot, Krutarth and Xiao, Boyuan and Han, Zherui and Guo, Ziqi and Xiong, Zixin and Ruan, Xiulin},
  title   = {Phonon local non-equilibrium at {Al}/{Si} interface from machine learning molecular dynamics},
  journal = {Journal of Applied Physics},
  year    = {2025},
  volume  = {137},
  number  = {11},
  pages   = {115301},
  doi     = {10.1063/5.0243641},
}

@article{Yang2023JAP,
  author  = {Yang, Hong-Ao and Cao, Bing-Yang},
  title   = {Mode-resolved phonon transmittance using lattice dynamics: Robust algorithm and statistical characteristics},
  journal = {Journal of Applied Physics},
  year    = {2023},
  volume  = {134},
  number  = {15},
  pages   = {155302},
  doi     = {10.1063/5.0171201},
}

@article{Haas2023NanoLett,
  author  = {Haas, Benedikt and Boland, Tara M. and Els{\"a}sser, Christian and Singh, Arunima K. and March, Katia and Barthel, Juri and Koch, Christoph T. and Rez, Peter},
  title   = {Atomic-Resolution Mapping of Localized Phonon Modes at Grain Boundaries},
  journal = {Nano Letters},
  year    = {2023},
  volume  = {23},
  number  = {13},
  pages   = {5975--5980},
  doi     = {10.1021/acs.nanolett.3c01089},
}

@article{Yang2017SciRep,
  author  = {Yang, Lina and Minnich, Austin J.},
  title   = {Thermal transport in nanocrystalline Si and SiGe by ab initio based Monte Carlo simulation},
  journal = {Scientific Reports},
  year    = {2017},
  volume  = {7},
  number  = {1},
  pages   = {44254},
  doi     = {10.1038/srep44254},
}

@article{Termentzidis2009PhysRevB,
  author  = {Termentzidis, Konstantinos and Chantrenne, Patrice and Keblinski, Pawel},
  title   = {Nonequilibrium molecular dynamics simulation of the in-plane thermal conductivity of superlattices with rough interfaces},
  journal = {Physical Review B},
  year    = {2009},
  volume  = {79},
  number  = {21},
  pages   = {214307},
  doi     = {10.1103/PhysRevB.79.214307},
}

@article{Bodapati2006PhysRevB,
  author  = {Bodapati, Arun and Schelling, Patrick K. and Phillpot, Simon R. and Keblinski, Pawel},
  title   = {Vibrations and thermal transport in nanocrystalline silicon},
  journal = {Physical Review B},
  year    = {2006},
  volume  = {74},
  number  = {24},
  pages   = {245207},
  doi     = {10.1103/PhysRevB.74.245207},
}

@article{Isotta2024AdvFunctMater,
  author  = {Isotta, Eleonora and Jiang, Shizhou and Bueno-Villoro, Ruben and Nagahiro, Ryohei and Maeda, Kosuke and Mattlat, Dominique Alexander and Odufisan, Alesanmi R. and Zevalkink, Alexandra and Shiomi, Junichiro and Zhang, Siyuan and Scheu, Christina and Snyder, G. Jeffrey and Balogun, Oluwaseyi},
  title   = {Heat Transport at Silicon Grain Boundaries},
  journal = {Advanced Functional Materials},
  year    = {2024},
  volume  = {34},
  number  = {40},
  pages   = {2405413},
  doi     = {10.1002/adfm.202405413},
}

@article{Wang2011NanoLett,
  author  = {Wang, Zhaojie and Alaniz, Joseph E. and Jang, Wanyoung and Garay, Javier E. and Dames, Chris},
  title   = {Thermal Conductivity of Nanocrystalline Silicon: Importance of Grain Size and Frequency-Dependent Mean Free Paths},
  journal = {Nano Letters},
  volume  = {11},
  number  = {6},
  pages   = {2206--2213},
  year    = {2011},
  doi     = {10.1021/nl1045395}
}

@article{Ju2013JAP,
  author  = {Ju, Sheng-Hong and Liang, Xin-Gang},
  title   = {Investigation on interfacial thermal resistance and phonon scattering at twist boundary of silicon},
  journal = {Journal of Applied Physics},
  volume  = {113},
  number  = {5},
  pages   = {053513},
  year    = {2013},
  doi     = {10.1063/1.4790178}
}

@article{Hua2017PhysRevB,
  author  = {Hua, Chengyun and Chen, Xiangwen and Ravichandran, Navaneetha K. and Minnich, Austin J.},
  title   = {Experimental metrology to obtain thermal phonon transmission coefficients at solid interfaces},
  journal = {Physical Review B},
  year    = {2017},
  volume  = {95},
  number  = {20},
  pages   = {205423},
  doi     = {10.1103/PhysRevB.95.205423},
}

@article{Corradini2025npjCompMater,
  author  = {Corradini, Andrea and Marini, Giovanni and Calandra, Matteo},
  title   = {Scalable machine learning approach to light induced order disorder phase transitions with ab initio accuracy},
  journal = {npj Computational Materials},
  year    = {2025},
  volume  = {11},
  number  = {1},
  pages   = {151},
  doi     = {10.1038/s41524-025-01614-5},
}

@article{Wang2023PhysRevB,
  author  = {Wang, Yanzhou and Fan, Zheyong and Qian, Ping and Caro, Miguel A. and Ala-Nissila, Tapio},
  title   = {Quantum-corrected thickness-dependent thermal conductivity in amorphous silicon predicted by machine learning molecular dynamics simulations},
  journal = {Physical Review B},
  year    = {2023},
  volume  = {107},
  number  = {5},
  pages   = {054303},
  doi     = {10.1103/PhysRevB.107.054303},
}

@article{Li2012PhysRevB,
  author  = {Li, Wu and Mingo, Natalio and Lindsay, L. and Broido, D. A. and Stewart, D. A. and Katcho, N. A.},
  title   = {Thermal conductivity of diamond nanowires from first principles},
  journal = {Physical Review B},
  year    = {2012},
  volume  = {85},
  number  = {19},
  pages   = {195436},
  doi     = {10.1103/PhysRevB.85.195436},
}

@article{Regner2013NatCommun,
  author  = {Regner, Keith T. and Sellan, Daniel P. and Su, Zonghui and Amon, Cristina H. and McGaughey, Alan J. H. and Malen, Jonathan A.},
  title   = {Broadband phonon mean free path contributions to thermal conductivity measured using frequency domain thermoreflectance},
  journal = {Nature Communications},
  volume  = {4},
  pages   = {1640},
  year    = {2013},
  doi     = {10.1038/ncomms2630},
  url     = {https://doi.org/10.1038/ncomms2630}
}

@article{Shanks1963PhysRev,
  author  = {Shanks, H. R. and Maycock, P. D. and Sidles, P. H. and Danielson, G. C.},
  title   = {Thermal Conductivity of Silicon from 300 to 1400 K},
  journal = {Physical Review},
  year    = {1963},
  volume  = {130},
  number  = {5},
  pages   = {1743--1748},
  doi     = {10.1103/PhysRev.130.1743},
}

@article{Glassbrenner1964PhysRev,
  author  = {Glassbrenner, C. J. and Slack, Glen A.},
  title   = {Thermal Conductivity of Silicon and Germanium from 3 K to the Melting Point},
  journal = {Physical Review},
  year    = {1964},
  volume  = {134},
  number  = {4A},
  pages   = {A1058--A1069},
  doi     = {10.1103/PhysRev.134.A1058},
}

@article{Johnson2013,
  title = {Direct Measurement of Room-Temperature Nondiffusive Thermal Transport Over Micron Distances in a Silicon Membrane},
  author = {Johnson, Jeremy A. and Maznev, A. A. and Cuffe, John and Eliason, Jeffrey K. and Minnich, Austin J. and Kehoe, Timothy and Torres, Clivia M. Sotomayor and Chen, Gang and Nelson, Keith A.},
  journal = {Phys. Rev. Lett.},
  volume = {110},
  issue = {2},
  pages = {025901},
  numpages = {5},
  year = {2013},
  month = {Jan},
  publisher = {American Physical Society},
  doi = {10.1103/PhysRevLett.110.025901},
  url = {https://link.aps.org/doi/10.1103/PhysRevLett.110.025901}
}

@article{Maire2022AdvFunctMater,
  author  = {Maire, Jeremie and Ch{\'a}vez-{\'A}ngel, Emigdio and Arregui, Guillermo and Colombano, Martin F. and Capuj, Nestor E. and Griol, Amadeu and Mart{\'\i}nez, Alejandro and Navarro-Urrios, Daniel and Ahopelto, Jouni and Sotomayor-Torres, Clivia M.},
  title   = {Thermal Properties of Nanocrystalline Silicon Nanobeams},
  journal = {Advanced Functional Materials},
  year    = {2022},
  volume  = {32},
  number  = {4},
  pages   = {2105767},
  doi     = {10.1002/adfm.202105767},
}

@article{Fujii2022CompMaterSci,
  author  = {Fujii, Susumu and Seko, Atsuto},
  title   = {Structure and lattice thermal conductivity of grain boundaries in silicon by using machine learning potential and molecular dynamics},
  journal = {Computational Materials Science},
  year    = {2022},
  volume  = {204},
  pages   = {111137},
  doi     = {10.1016/j.commatsci.2021.111137},
}

@article{Deringer2019AdvMater,
  author  = {Deringer, Volker L. and Caro, Miguel A. and Cs{\'a}nyi, G{\'a}bor},
  title   = {Machine Learning Interatomic Potentials as Emerging Tools for Materials Science},
  journal = {Advanced Materials},
  year    = {2019},
  volume  = {31},
  number  = {46},
  pages   = {1902765},
  doi     = {10.1002/adma.201902765},
}

@article{Protik2022,
  author  = {Protik, Nakib H. and Li, Chunhua and Pruneda, Miguel and Broido, David and Ordej{\'o}n, Pablo},
  title   = {The {elphbolt} ab initio solver for the coupled electron--phonon Boltzmann transport equations},
  journal = {npj Computational Materials},
  year    = {2022},
  volume  = {8},
  number  = {1},
  pages   = {28},
  doi     = {10.1038/s41524-022-00710-0},
}

@article{PhysRevB.102.195412,
  title = {Quantum mechanical modeling of anharmonic phonon-phonon scattering in nanostructures},
  author = {Guo, Yangyu and Bescond, Marc and Zhang, Zhongwei and Luisier, Mathieu and Nomura, Masahiro and Volz, Sebastian},
  journal = {Phys. Rev. B},
  volume = {102},
  issue = {19},
  pages = {195412},
  numpages = {24},
  year = {2020},
  month = {Nov},
  publisher = {American Physical Society},
  doi = {10.1103/PhysRevB.102.195412},
  url = {https://link.aps.org/doi/10.1103/PhysRevB.102.195412}
}

@article{Li2019FrontPhys,
  author  = {Li, Chen and Tian, Zhiting},
  title   = {Phonon Transmission Across Silicon Grain Boundaries by Atomistic Green's Function Method},
  journal = {Frontiers in Physics},
  volume  = {7},
  pages   = {3},
  year    = {2019},
  doi     = {10.3389/fphy.2019.00003}
}

@article{Nose1984JCP,
  author  = {Nos{\'e}, Shuichi},
  title   = {A unified formulation of the constant temperature molecular dynamics methods},
  journal = {The Journal of Chemical Physics},
  volume  = {81},
  number  = {1},
  pages   = {511--519},
  year    = {1984},
  doi     = {10.1063/1.447334}
}

@article{Hoover1985PRA,
  title = {Canonical dynamics: Equilibrium phase-space distributions},
  author = {Hoover, William G.},
  journal = {Phys. Rev. A},
  volume = {31},
  issue = {3},
  pages = {1695--1697},
  numpages = {0},
  year = {1985},
  month = {Mar},
  publisher = {American Physical Society},
  doi = {10.1103/PhysRevA.31.1695},
  url = {https://link.aps.org/doi/10.1103/PhysRevA.31.1695}
}

@article{Dudde2025MTP,
  author  = {Dudde, Katharina and Elhajhasan, Mahmoud and Würsch, Guillaume and Themann, Julian and Lierath, Jana and Paul, Dwaipayan and Protik, Nakib H. and Romano, Giuseppe and Callsen, Gordon},
  title   = {Phonon mean free path spectroscopy by Raman thermometry},
  journal = {Materials Today Physics},
  year    = {2025},
  volume  = {57},
  pages   = {101784},
  doi     = {10.1016/j.mtphys.2025.101784},
}

@article{PhysRevE.83.056706,
  title = {Kapitza resistance in the lattice Boltzmann-Peierls-Callaway equation for multiphase phonon gases},
  author = {Lee, Jonghoon and Roy, Ajit K. and Farmer, Barry L.},
  journal = {Phys. Rev. E},
  volume = {83},
  issue = {5},
  pages = {056706},
  numpages = {11},
  year = {2011},
  month = {May},
  publisher = {American Physical Society},
  doi = {10.1103/PhysRevE.83.056706},
  url = {https://link.aps.org/doi/10.1103/PhysRevE.83.056706}
}

@article{Maiti1997SSC,
  author  = {Maiti, A. and Mahan, G. D. and Pantelides, S. T.},
  title   = {Dynamical simulations of nonequilibrium processes — Heat flow and the Kapitza resistance across grain boundaries},
  journal = {Solid State Communications},
  volume  = {102},
  number  = {7},
  pages   = {517--521},
  year    = {1997},
  doi     = {10.1016/S0038-1098(97)00049-5}
}

@article{Wang2026MTP,
  author  = {Wang, Weitao and Wu, Xin and Wu, Yunhui and Volz, Sebastian and Takagi, Takeshi and Nomura, Masahiro},
  title   = {Interfacial thermal transport analysis using machine learning potential at AlN/Cu van der Waals interface},
  journal = {Materials Today Physics},
  volume  = {62},
  pages   = {102057},
  year    = {2026},
  doi     = {10.1016/j.mtphys.2026.102057}
}

@article{Tadano2014JPCM,
  author  = {Tadano, T. and Gohda, Y. and Tsuneyuki, S.},
  title   = {Anharmonic force constants extracted from first-principles molecular dynamics: applications to heat transfer simulations},
  journal = {Journal of Physics: Condensed Matter},
  volume  = {26},
  number  = {22},
  pages   = {225402},
  year    = {2014},
  doi     = {10.1088/0953-8984/26/22/225402}
}

@article{Cheng2021NatCommun,
  author  = {Cheng, Zhe and Li, Ruiyang and Yan, Xingxu and Jernigan, Glenn and Shi, Jingjing and Liao, Michael E. and Hines, Nicholas J. and Gadre, Chaitanya A. and Idrobo, Juan Carlos and Lee, Eungkyu and Hobart, Karl D. and Goorsky, Mark S. and Pan, Xiaoqing and Luo, Tengfei and Graham, Samuel},
  title   = {Experimental observation of localized interfacial phonon modes},
  journal = {Nature Communications},
  volume  = {12},
  number  = {1},
  pages   = {6901},
  year    = {2021},
  doi     = {10.1038/s41467-021-27250-3}
}

@article{PhysRevB.35.9120,
  title = {Phase diagram of silicon by molecular dynamics},
  author = {Broughton, J. Q. and Li, X. P.},
  journal = {Phys. Rev. B},
  volume = {35},
  issue = {17},
  pages = {9120--9127},
  numpages = {0},
  year = {1987},
  month = {Jun},
  publisher = {American Physical Society},
  doi = {10.1103/PhysRevB.35.9120},
  url = {https://link.aps.org/doi/10.1103/PhysRevB.35.9120}
}

@article{Thompson2022LAMMPS,
  author  = {Thompson, Aidan P. and Aktulga, H. Metin and Berger, Richard and Bolintineanu, Dan S. and Brown, W. Michael and Crozier, Paul S. and in 't Veld, Pieter J. and Kohlmeyer, Axel and Moore, Stan G. and Nguyen, Trung Dac and Shan, Ray and Stevens, Mark J. and Tranchida, Julien and Trott, Christian and Plimpton, Steven J.},
  title   = {LAMMPS - A flexible simulation tool for particle-based materials modeling at the atomic, meso, and continuum scales},
  journal = {Computer Physics Communications},
  volume  = {271},
  pages   = {108171},
  year    = {2022},
  doi     = {10.1016/j.cpc.2021.108171}
}

@article{Hirel2015Atomsk,
  author  = {Hirel, Pierre},
  title   = {Atomsk: A tool for manipulating and converting atomic data files},
  journal = {Computer Physics Communications},
  volume  = {197},
  pages   = {212--219},
  year    = {2015},
  doi     = {10.1016/j.cpc.2015.07.012}
}

@article{HjorthLarsen2017ASE,
  author  = {Hjorth Larsen, Ask and Mortensen, Jens J{\o}rgen and Blomqvist, Jakob and others},
  title   = {The Atomic Simulation Environment: A Python library for working with atoms},
  journal = {Journal of Physics: Condensed Matter},
  volume  = {29},
  number  = {27},
  pages   = {273002},
  year    = {2017},
  doi     = {10.1088/1361-648X/aa680e}
}

@article{phonopy-phono3py-JPCM,
  author  = {Togo, Atsushi and Chaput, Laurent and Tadano, Terumasa and Tanaka, Isao},
  title   = {Implementation strategies in phonopy and phono3py},
  journal = {J. Phys. Condens. Matter},
  volume  = {35},
  number  = {35},
  pages   = {353001},
  year    = {2023},
  doi     = {10.1088/1361-648X/acd831}
}

@article{phonopy-phono3py-JPSJ,
  author  = {Togo, Atsushi},
  title   = {First-principles Phonon Calculations with Phonopy and Phono3py},
  journal = {J. Phys. Soc. Jpn.},
  volume  = {92},
  number  = {1},
  pages   = {012001},
  year    = {2023},
  doi     = {10.7566/JPSJ.92.012001}
}

@article{Yang2020,
  author  = {Yang, Cheng and Wei, Xinrui and Sheng, Jiteng and Wu, Haibin},
  title   = {Phonon heat transport in cavity-mediated optomechanical nanoresonators},
  journal = {Nature Communications},
  volume  = {11},
  number  = {1},
  pages   = {4656},
  year    = {2020},
  doi     = {10.1038/s41467-020-18426-4}
}

@article{Morell2019MoSe2,
  author  = {Morell, Nicolas and Tepsic, Slaven and Reserbat-Plantey, Antoine and Cepellotti, Andrea and Manca, Marco and Epstein, Itai and Isacsson, Andreas and Marie, Xavier and Mauri, Francesco and Bachtold, Adrian},
  title   = {Optomechanical Measurement of Thermal Transport in Two-Dimensional {MoSe2} Lattices},
  journal = {Nano Letters},
  year    = {2019},
  volume  = {19},
  number  = {5},
  pages   = {3143--3150},
  doi     = {10.1021/acs.nanolett.9b00560}
}

@article{Jia2026,
  author  = {Jia, Hao and Ye, Fan and Feng, Philip X.-L.},
  title   = {Gigahertz Multimode Vibrations in Graphene and MoS$_2$ Nanomechanical Resonators at Room Temperature},
  journal = {Science Advances},
  year    = {2026},
  volume  = {12},
  number  = {7},
  pages   = {eads5668},
  doi     = {10.1126/sciadv.ads5668}
}

\clearpage

\section*{Supplementary Material}

\renewcommand{\thefigure}{S\arabic{figure}}
\renewcommand{\thetable}{S\arabic{table}}
\renewcommand{\theequation}{S\arabic{equation}}

\setcounter{figure}{0}
\setcounter{table}{0}
\setcounter{equation}{0}

This Supplementary Material details the computational methodology, validation procedures, and additional analyzes supporting the main manuscript. It outlines the bulk and grain-boundary workflows, including structure preparation, equilibration protocols, finite-temperature sampling, and the extraction of thermodynamic and transport properties. Convergence tests and key parameter choices are also provided to ensure transparency and reproducibility of the results.

\setcounter{section}{0}
\renewcommand{\thesection}{S\arabic{section}}
\section{Bulk Silicon Workflow}

\begin{figure}[!htbp]
    \centering
    \includegraphics[width=0.8\linewidth]{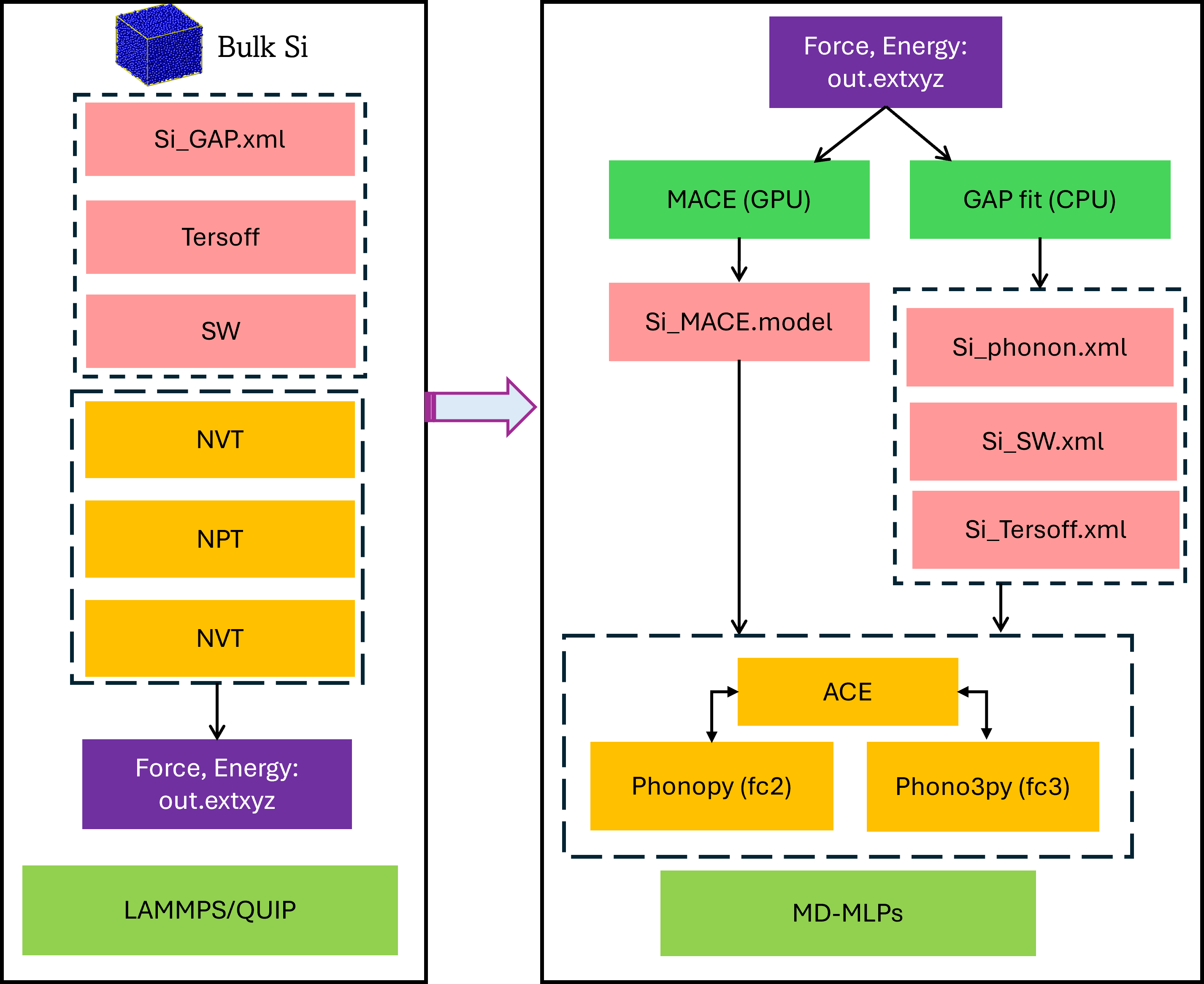}
    \caption{Schematic of the bulk silicon workflow for machine-learning potential training and force-constant extraction.}
    \label{fig:bulk_workflow}
\end{figure}

\begin{figure}[!htbp]
    \centering
    \includegraphics[width=0.7\linewidth]{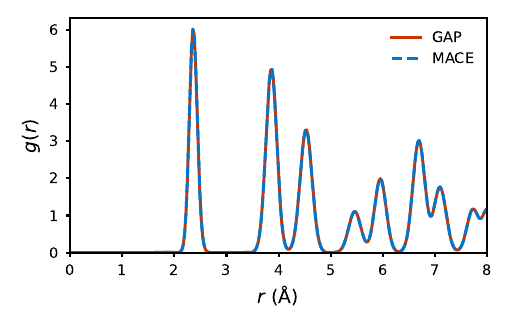}
    \caption{Radial distribution function $g(r)$ of bulk silicon predicted by the GAP and MACE models.}
    \label{fig:rdf_bulk}
\end{figure}

The bulk workflow is summarized schematically in Figure~\ref{fig:bulk_workflow}. A periodic $3 \times 3 \times 3$ silicon supercell (216 atoms) was used to generate reference configurations. After structural relaxation, finite-temperature snapshots were collected at 300~K using short equilibrium simulations with a 1~fs time step. Configurations were stored together with the corresponding total energies and atomic forces in \texttt{extxyz} format. This single dataset served as the common training input for all interatomic models, with GAP~\cite{Bartok2013PhysRevB,Bartok2018PRX,Caro2019PhysRevB} and MACE~\cite{Batatia2022NeurIPS,Kovacs2023JCP} trained on the same reference file to ensure a consistent comparison, while Stillinger-Weber and Tersoff were fitted to the identical data. The trained models were subsequently interfaced with ASE for lattice-dynamical calculations~\cite{HjorthLarsen2017ASE}. Harmonic and anharmonic force constants were obtained from finite-displacement calculations on the relaxed supercell using displacement amplitudes of 0.01~\AA\ (second order) and 0.03~\AA\ (third order). The force-constant tensors were reconstructed using Phonopy and Phono3py~\cite{phonopy-phono3py-JPCM} for subsequent vibrational and thermal transport analysis. Figure~\ref{fig:rdf_bulk} shows that the radial distribution functions predicted by GAP and MACE are nearly indistinguishable across the full range of interatomic distances. The first sharp peak at approximately 2.35~\AA\ corresponds to the nearest-neighbor Si-Si bond length in the diamond cubic structure, while subsequent peaks reproduce higher coordination shells with consistent positions and amplitudes. The excellent agreement indicates that both machine-learning potentials accurately capture the equilibrium structural correlations and preserve the underlying crystalline order of bulk silicon.

\section{Grain-Boundary Workflow and Dataset Construction}

\begin{figure}[!htbp]
    \centering
    \includegraphics[width=0.8\linewidth]{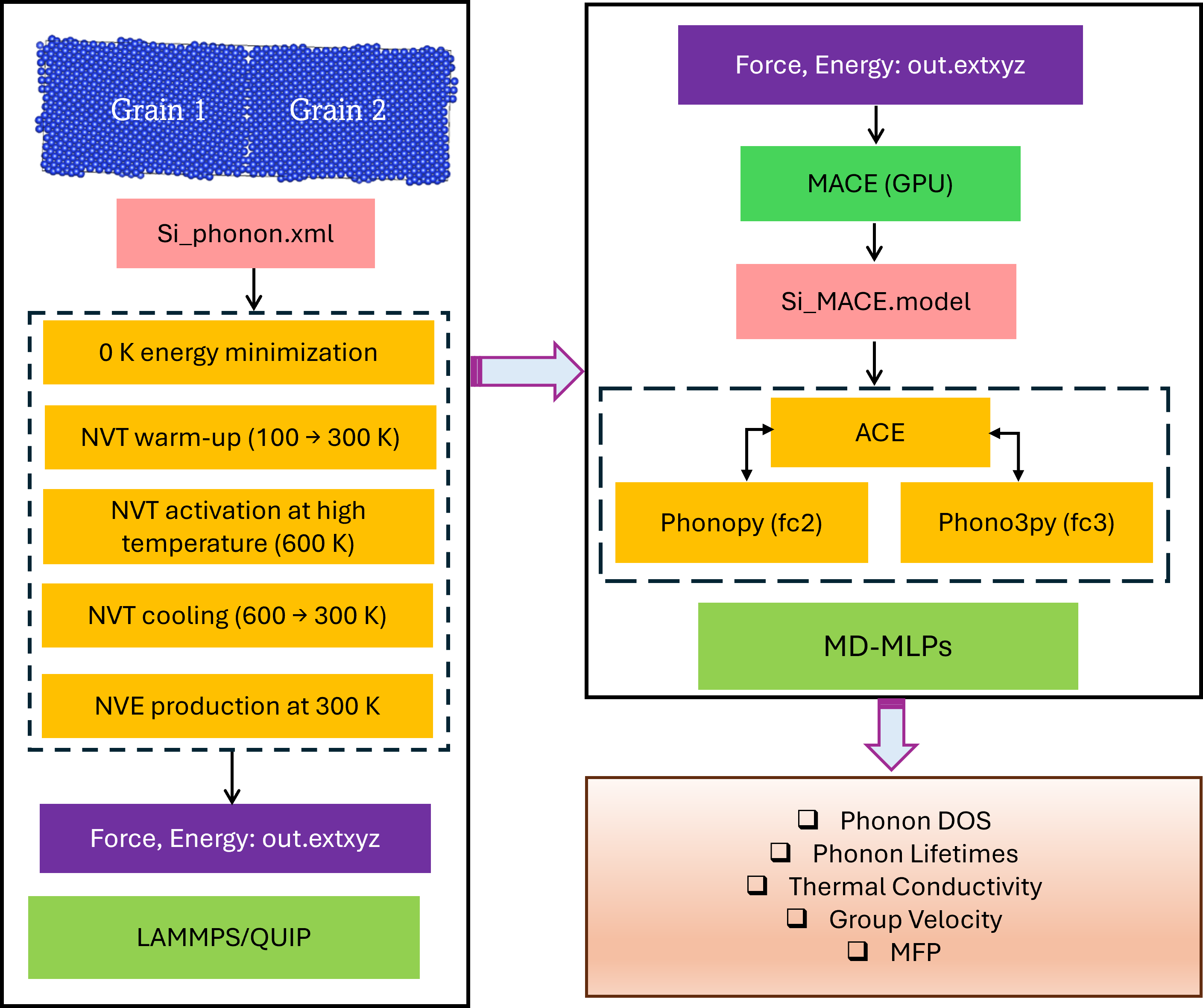}
    \caption{Schematic of the grain-boundary workflow.}
    \label{fig:gb_workflow}
\end{figure}

\begin{figure}[!htbp]
    \centering
    \includegraphics[width=0.7\linewidth]{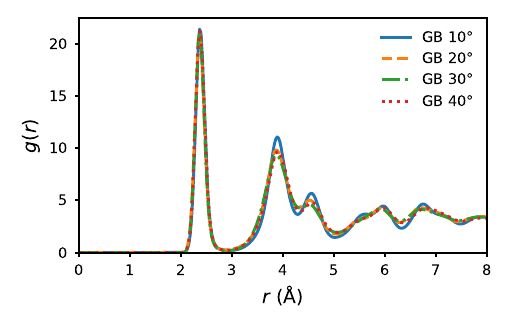}
    \caption{Radial distribution function $g(r)$ for grain boundaries with misorientation angles of $10^\circ$, $20^\circ$, $30^\circ$, and $40^\circ$.}
    \label{fig:GB_RDF}
\end{figure}

\begin{figure}[!htbp]
    \centering
    \begin{subfigure}{0.48\linewidth}
        \centering
        \includegraphics[width=\linewidth]{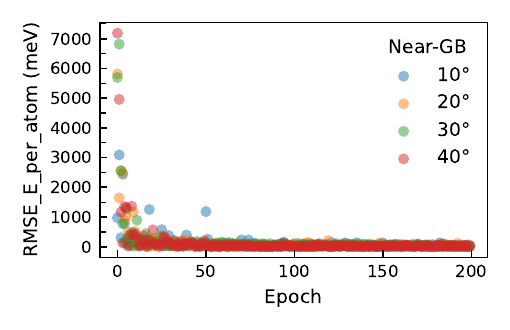}
        \caption{}
        \label{fig:rmse_bulklike}
    \end{subfigure}
    \hfill
    \begin{subfigure}{0.48\linewidth}
        \centering
        \includegraphics[width=\linewidth]{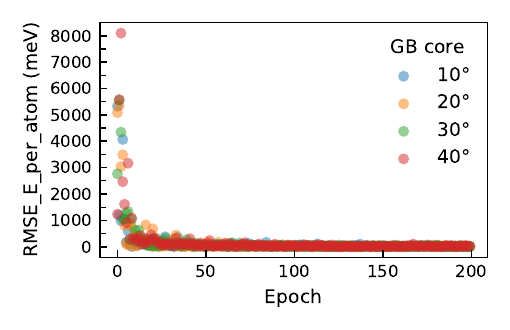}
        \caption{}
        \label{fig:rmse_gbcore}
    \end{subfigure}
    \caption{Training convergence of the energy error (RMSE$_{E}$/atom) for models fitted to datasets with different misorientation angles: (a) Near-GB region configurations (atoms selected away from the interface) and (b) grain-boundary core configurations.}
    \label{fig:training_rmse_angles}
\end{figure}

\begin{figure}[t]
    \centering
    \begin{subfigure}{0.48\linewidth}
        \centering
        \includegraphics[width=\linewidth]{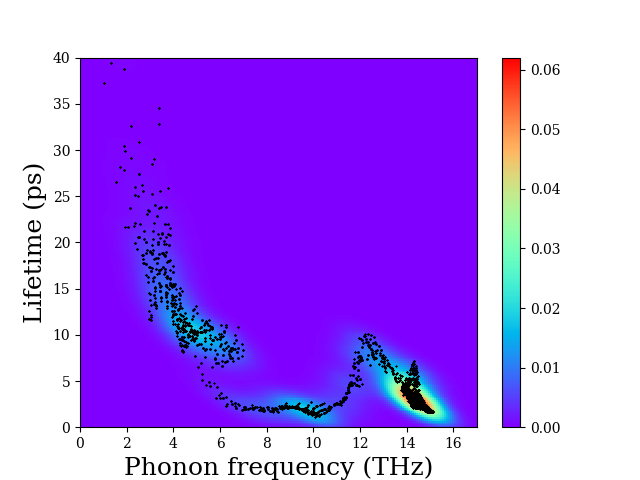}
        \caption{$10^\circ$}
        \label{fig:bulk_lifetime_10}
    \end{subfigure}
    \hfill
    \begin{subfigure}{0.48\linewidth}
        \centering
        \includegraphics[width=\linewidth]{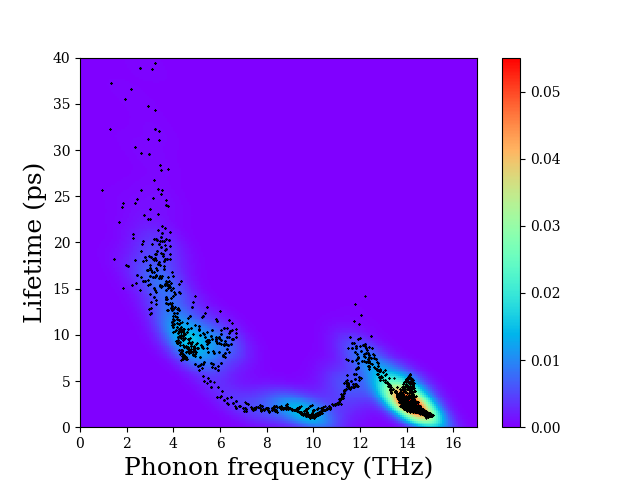}
        \caption{$20^\circ$}
        \label{fig:bulk_lifetime_20}
    \end{subfigure}
    \hfill
    \begin{subfigure}{0.48\linewidth}
        \centering
        \includegraphics[width=\linewidth]{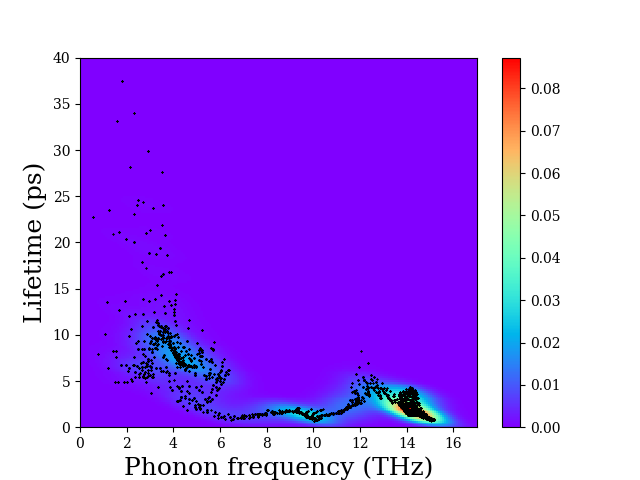}
        \caption{$30^\circ$}
        \label{fig:bulk_lifetime_30}
    \end{subfigure}
    \hfill
    \begin{subfigure}{0.48\linewidth}
        \centering
        \includegraphics[width=\linewidth]{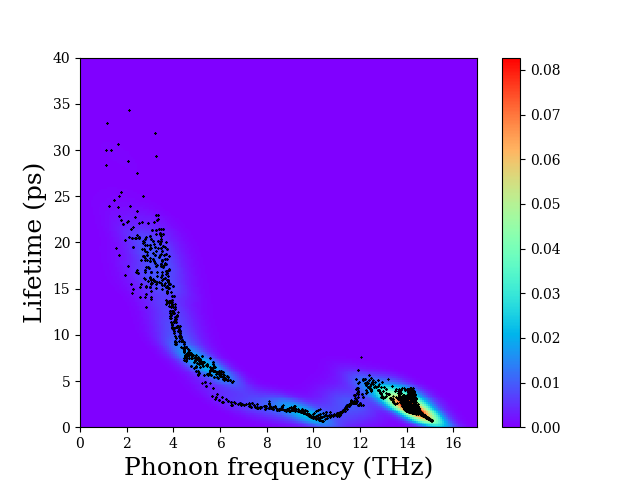}
        \caption{$40^\circ$}
        \label{fig:bulk_lifetime_40}
    \end{subfigure}

    \caption{Phonon lifetimes in the near-GB region for different misorientation angles: (a) $10^\circ$, (b) $20^\circ$, (c) $30^\circ$, and (d) $40^\circ$ at 300~K.}
    \label{fig:bulk_lifetime}
\end{figure}

This section provides implementation-specific details of the grain-boundary computational framework, including bicrystal preparation, dataset generation, and lattice-dynamical calculations within the interfacial region. The description focuses on reproducibility and complements the methodological overview presented in the main text. As illustrated in Figure~\ref{fig:gb_workflow}, relaxed bicrystal configurations were subjected to a staged thermal protocol to sample representative interfacial atomic environments. Energies and forces extracted from these trajectories were written in \texttt{extxyz} format and used to train a MACE interatomic potential specifically for the grain-boundary structures. The trained model was then interfaced with ASE to perform finite-displacement calculations, from which second- and third-order force constants were reconstructed for subsequent vibrational and thermal-transport analyzes within the interfacial region. Figure~\ref{fig:GB_RDF} compares the radial distribution functions (RDFs) for the different grain-boundary misorientations. The RDFs were computed from the selected GB-core region, comprising approximately 579 atoms within the interval $24 \le x \le 37$~\text{\AA}. This region includes not only atoms located directly at the grain boundary but also neighboring atomic layers surrounding the interface in order to provide sufficient statistical sampling. Consequently, the RDF retains a significant contribution from locally ordered crystalline environments, which explains the relatively sharp first-neighbor peak observed for all misorientation angles. Differences emerge primarily in the second- and third-neighbor coordination shells, where slight peak broadening and reductions in peak intensity are observed with increasing misorientation angle, reflecting moderate distortions in the medium-range atomic order. Overall, the RDF profiles indicate increased disorder relative to bulk silicon while retaining a substantial degree of crystalline character within the grain-boundary region. All models were trained using the same MACE architecture and identical training settings for both the near-GB and grain-boundary datasets, including a cutoff radius of 5.5~\AA, a batch size of one configuration, a maximum of 200 training epochs, and an exponential moving average (EMA) of the model parameters with a decay factor of 0.99 to enhance numerical stability. Consequently, any differences in convergence behavior originate from the structural characteristics of the datasets rather than changes in the training procedure. Figure~\ref{fig:training_rmse_angles}(a) shows the evolution of the energy root-mean-square error per atom (RMSE$_E$/atom) for configurations extracted from near-GB region away from the interface. The error decreases rapidly during the early epochs and approaches a stable plateau, indicating efficient optimization and consistent learning across all misorientation angles. Figure~\ref{fig:training_rmse_angles}(b) presents the corresponding behavior for grain-boundary core configurations. The initial RMSE values are larger, and the convergence exhibits slightly broader fluctuations compared to the near-GB case, reflecting the greater diversity of local atomic environments present at the interface. Despite this increased structural complexity, the models reach comparable final error levels within the same training window, demonstrating robust performance under a unified training setup. Figure~\ref{fig:bulk_lifetime} shows that the phonon lifetime distributions in the near-GB region remain largely unchanged with misorientation angle and closely resemble those of pristine bulk silicon, confirming that vibrational properties are effectively recovered far from the grain boundary. A slight reduction in lifetimes is observed for the $40^\circ$ case, which may indicate a weak residual influence of interfacial disorder extending into the near-GB region.

\section{Machine-learning non-equilibrium molecular dynamics workflow}

\begin{figure}[!htbp]
    \centering
    \includegraphics[width=0.8\linewidth]{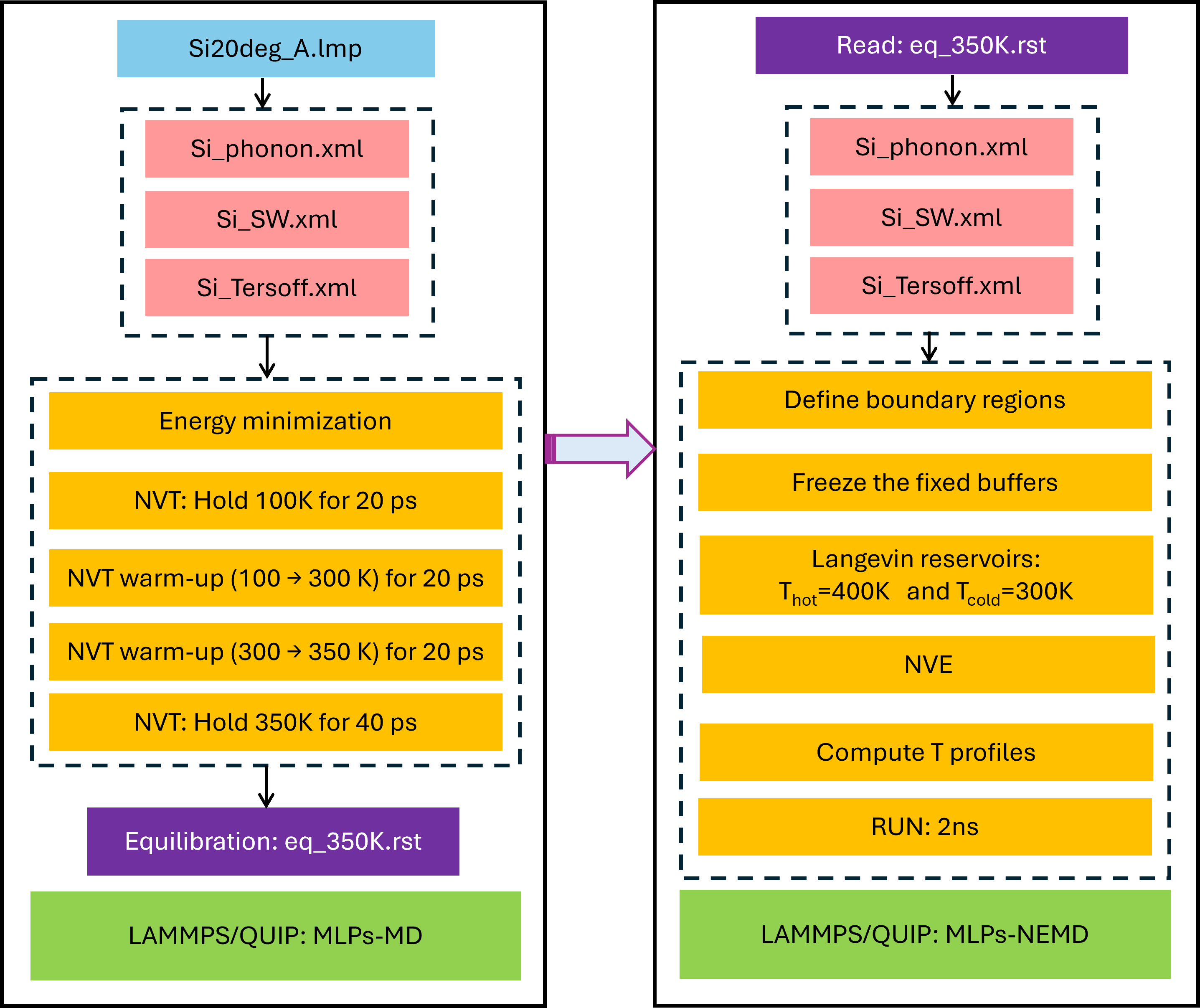}
    \caption{Workflow of the non-equilibrium molecular dynamics (NEMD) simulations.}
    \label{fig:nemd_workflow}
\end{figure}

\begin{figure}[t]
    \centering
    \begin{subfigure}{0.32\linewidth}
        \centering
        \includegraphics[width=\linewidth]{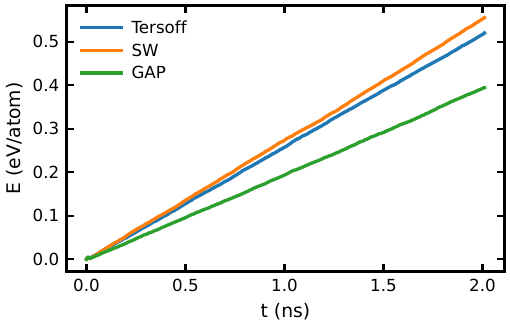}
        \caption{}
        \label{fig:nemd_energy_A1}
    \end{subfigure}
    \hfill
    \begin{subfigure}{0.32\linewidth}
        \centering
        \includegraphics[width=\linewidth]{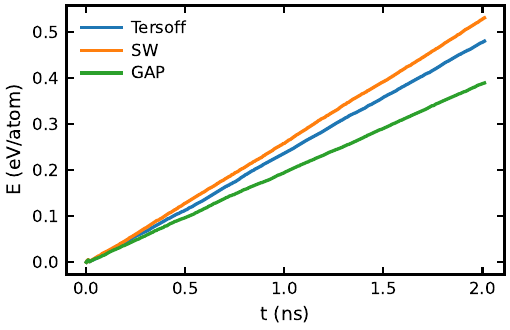}
        \caption{}
        \label{fig:nemd_energy_A2}
    \end{subfigure}
    \hfill
    \begin{subfigure}{0.32\linewidth}
        \centering
        \includegraphics[width=\linewidth]{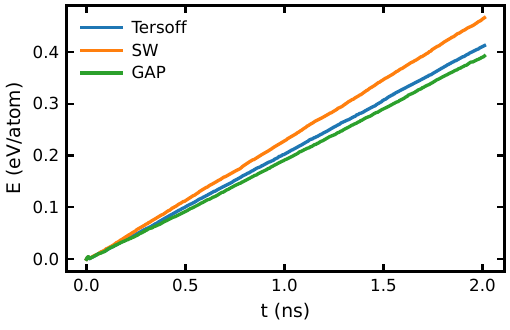}
        \caption{}
        \label{fig:nemd_energy_A3}
    \end{subfigure}

    \caption{Non-equilibrium molecular dynamics simulations. Time evolution of the average potential energy per atom for different roughness amplitudes: (a) $A = 1$~\text{\AA}, (b) $A = 2$~\text{\AA}, and (c) $A = 3$~\text{\AA}.}
    \label{fig:nemd_energy}
\end{figure}

\begin{figure}[t]
    \centering
    \begin{subfigure}{0.48\linewidth}
        \centering
        \includegraphics[width=\linewidth]{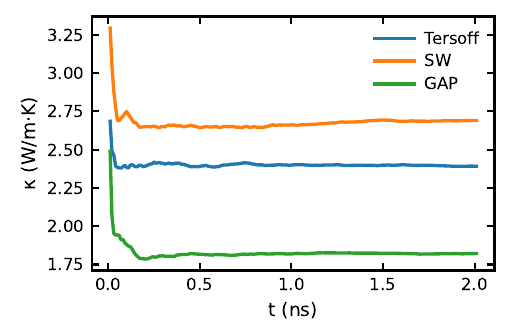}
        \caption{}
        \label{fig:nemd_kappa_A1}
    \end{subfigure}
    \hfill
    \begin{subfigure}{0.48\linewidth}
        \centering
        \includegraphics[width=\linewidth]{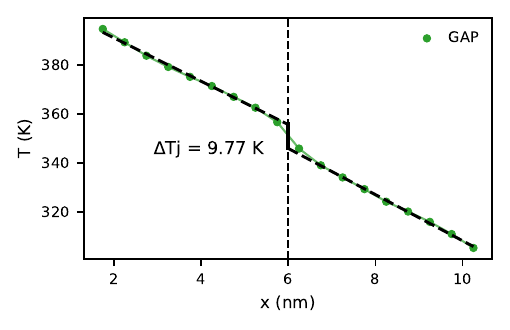}
        \caption{}
        \label{fig:Tprofile_gap_A1}
    \end{subfigure}

    \vspace{0.6em}

    \begin{subfigure}{0.48\linewidth}
        \centering
        \includegraphics[width=\linewidth]{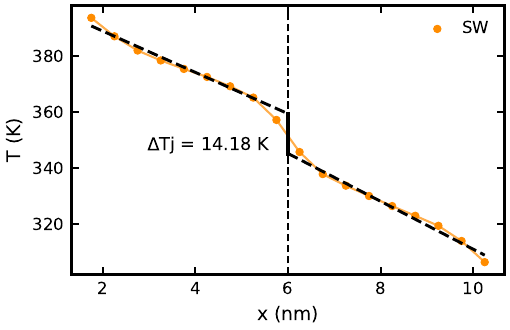}
        \caption{}
        \label{fig:Tprofile_sw_A1}
    \end{subfigure}
    \hfill
    \begin{subfigure}{0.48\linewidth}
        \centering
        \includegraphics[width=\linewidth]{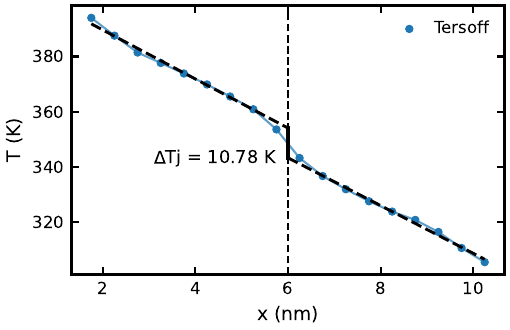}
        \caption{}
        \label{fig:Tprofile_ter_A1}
    \end{subfigure}

\caption{Non-equilibrium molecular dynamics simulations for $A = 1$~\text{\AA}. (a) Time evolution of $\kappa$. (b--d) Temperature profiles obtained with GAP, SW, and Tersoff, respectively.}
\label{fig:nemd_results_A1}
\end{figure}

\begin{figure}[t]
    \centering
    \begin{subfigure}{0.48\linewidth}
        \centering
        \includegraphics[width=\linewidth]{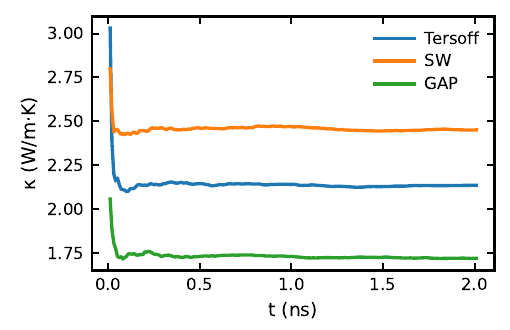}
        \caption{}
        \label{fig:nemd_kappa_A2}
    \end{subfigure}
    \hfill
    \begin{subfigure}{0.48\linewidth}
        \centering
        \includegraphics[width=\linewidth]{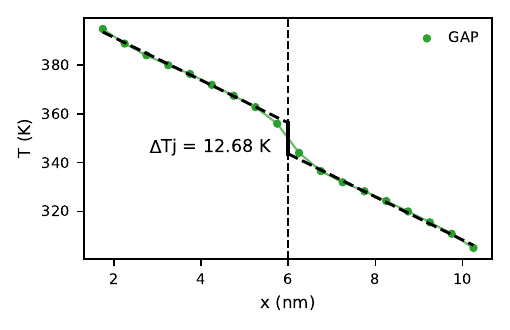}
        \caption{}
        \label{fig:Tprofile_gap_A2}
    \end{subfigure}

    \vspace{0.6em}

    \begin{subfigure}{0.48\linewidth}
        \centering
        \includegraphics[width=\linewidth]{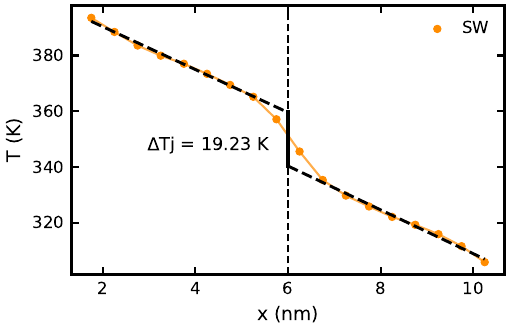}
        \caption{}
        \label{fig:Tprofile_sw_A2}
    \end{subfigure}
    \hfill
    \begin{subfigure}{0.48\linewidth}
        \centering
        \includegraphics[width=\linewidth]{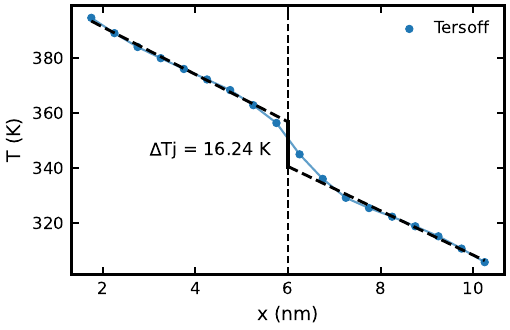}
        \caption{}
        \label{fig:Tprofile_ter_A2}
    \end{subfigure}

\caption{Non-equilibrium molecular dynamics simulations for $A = 2$~\text{\AA}. (a) Time evolution of $\kappa$. (b--d) Temperature profiles obtained with GAP, SW, and Tersoff, respectively.}

\label{fig:nemd_results_A2}
\end{figure}

This section provides additional details of the NEMD simulation procedure used to compute interfacial thermal transport. The detailed NEMD implementation is summarized in Fig.~\ref{fig:nemd_workflow}.  All simulations were performed with a timestep of 1~fs. Following energy minimization, the rough bicrystal structures were equilibrated under NVT dynamics using a staged protocol consisting of a 20~ps hold at 100~K, successive 20~ps temperature ramps to 300~K and 350~K, and a final 40~ps hold at 350~K. The equilibrated state was written to a restart file and used directly for the transport simulations.
In the production stage, 5~\AA\ fixed buffer layers were applied at both ends of the simulation cell along the transport direction, adjacent to 15~\AA\ Langevin regions maintained at 400~K and 300~K. The Langevin damping parameter was set to 0.5~ps, and energy exchange with the thermostats was recorded to compute the heat flux.  All non-fixed atoms were integrated in the microcanonical ensemble. A 10~ps settling run preceded the main 2~ns production simulation. Temperature profiles were obtained by binning the mobile atoms along the transport direction with a spatial resolution of $\Delta x = 5$~\AA\ and accumulating a running time average of the kinetic temperature in each bin.
To assess the influence of interfacial geometry, three roughness amplitudes, $A = 1$~\AA\ , $A = 2$~\AA\ and $A = 3$~\AA, were considered. Starting from the symmetric bicrystal structure, a sinusoidal displacement field was applied locally within a slab centered at the nominal grain-boundary plane, $x_{\mathrm{GB}} = 60$~\AA. Atoms satisfying $|x - x_{\mathrm{GB}}| \le 6$~\AA\ were displaced along the transport direction according to
\begin{equation}
\Delta x(y) = A \sin\!\left(\frac{2\pi y}{L}\right),
\end{equation}
where the wavelength was fixed at $L = 12$~\AA. The perturbation was confined to this 12~\AA-thick interfacial region, leaving the adjoining crystalline bulk regions unmodified. The resulting configurations were written to data files and used directly as initial structures for the NEMD production simulations under otherwise identical conditions. To assess the stability of the NEMD simulations for different interface geometries, the time evolution of the effective thermal conductivity $\kappa(t)$ was monitored. Figure~\ref{fig:nemd_energy} shows the time evolution of the average potential energy per atom during the NEMD simulations for roughness amplitudes of (a) $A = 1$~\text{\AA}, (b) $A = 2$~\text{\AA}, and (c) $A = 3$~\text{\AA}. The similar behavior observed across all roughness amplitudes indicates that the energy injection process is consistent, while differences between the interatomic potentials are reflected in the relative magnitude of the energy evolution. Figures~\ref{fig:nemd_results_A1} and \ref{fig:nemd_results_A2} show the corresponding $\kappa(t)$ curves for roughness amplitudes $A=1$~\text{\AA} and $A=2$~\text{\AA}. In both cases, $\kappa(t)$ evolves toward well-defined values with moderate fluctuations, reflecting the finite size of the system. The final values are consistent with those reported in the main manuscript. The magnitude of the fluctuations is comparable across the different interatomic potentials, indicating similar levels of statistical uncertainty in the extracted transport properties. The temperature profiles for $A=1$~\text{\AA}, shown in Figure~\ref{fig:nemd_results_A1}(b--d), exhibit a relatively small temperature discontinuity, indicating weaker phonon scattering and more efficient heat transmission across the boundary. In addition, the temperature gradients in the bulk-like regions remain approximately linear, confirming that heat transport away from the interface is not significantly perturbed. Minor deviations from linearity near the thermostat regions are observed, which can be attributed to local non-equilibrium effects introduced by the Langevin reservoirs.

\clearpage
\section*{Supplementary Figures}

\end{document}